%% file: main.tex
\documentclass[12pt]{article}

\usepackage[bf,small]{caption}
\usepackage[export]{adjustbox}
\usepackage[sort&compress,square,numbers]{natbib}
\usepackage[title]{appendix}
\usepackage[top=2.5cm,left=2.7cm,right=2.7cm,bottom=3.2cm]{geometry}
\usepackage{amsmath,amsthm,amssymb,setspace,enumitem,epsfig,titlesec,verbatim,color,array,multirow,comment,graphicx}
\usepackage{amssymb,amsfonts,amsmath,amsthm}
\usepackage{authblk}
\usepackage{bbm}
\usepackage{fullpage}
\usepackage{graphbox}
\usepackage{graphicx, xcolor}
\usepackage{graphicx}
\usepackage{hyperref}
\usepackage[capitalize]{cleveref}
\usepackage{lipsum} 
\usepackage{multicol}
\usepackage{subcaption}
\usepackage{times}
\usepackage{nicefrac}
\usepackage{multirow}

\let\leq=\leqslant 
\let\geq=\geqslant %
\let\ge=\geqslant %

\makeatletter
\def\blfootnote{\gdef\@thefnmark{}\@footnotetext}
\makeatother

\smallskip

\titleformat{\section}{\sffamily \fontsize{12}{12}\bfseries}{\thesection}{1em}{}
\titleformat{\subsection}{\sffamily \fontsize{10}{10.5}\bfseries}{\thesubsection}{1em}{}

\newcommand{\SI}{{\bf SI}}

\newcommand{\SIModel}{Section~1}
\newcommand{\SIComplete}{Section~2}
\newcommand{\SICycle}{Section~3}

\newcommand{\SIStar}{Section~5}

\newcommand{\SIDoubleStar}{Section~7}
\newcommand{\SIBarbell}{Section~8}
\newcommand{\SIGeneral}{Section~9}
\newcommand{\SIDirected}{Section~10}
\newcommand{\SIExtendedDataFigures}{Section~11}

\title{\bfseries\sffamily \LARGE Maintaining diversity in structured populations}
\date{}
\author[a,b]{David A. Brewster}
\author[c]{Jakub Svoboda}
\author[d]{Dylan Roscow}
\author[c]{Krishnendu Chatterjee}
\author[e]{Josef Tkadlec}
\author[f,g]{Martin A. Nowak}

\affil[a]{John A. Paulson School of Engineering and Applied Sciences, Harvard University, Boston, MA~02134,~USA}
\affil[b]{Department of Molecular and Cellular Biology, Harvard University, Cambridge, MA~02138,~USA}
\affil[c]{Institute of Science and Technology Austria, Klosterneuburg, Austria}
\affil[d]{Department of Mathematics, University of Illinois at Urbana-Champaign, Urbana, IL~61801,~USA}
\affil[e]{Computer Science Institute, Charles University, Prague, Czech Republic}
\affil[f]{Department of Organismic and Evolutionary Biology, Harvard University, Cambridge MA~02138,~USA}
\affil[g]{Department of Mathematics, Harvard University, Cambridge, MA 02138, USA}

\begin{document}
\maketitle
\onehalfspacing

~\\[0.4cm]

\noindent
{\bf
We examine population structures for their ability to maintain diversity in neutral evolution.
We use the general framework of evolutionary graph theory and consider birth-death (bd) and death-birth (db) updating.
The population is of size $N$.
Initially all individuals represent different types.
The basic question is: what is the time $T_N$ until one type takes over the population?
This time is known as consensus time in computer science and as total coalescent time in evolutionary biology.
For the complete graph, it is known that $T_N$ is quadratic in $N$ for db and bd.
For the cycle, we prove that $T_N$ is cubic in $N$ for db and bd.
For the star, we prove that $T_N$ is cubic for bd and quasilinear ($N\log N$) for db.
For the double star, we show that $T_N$ is quartic for bd. We derive upper and lower bounds for all undirected graphs for bd and db.
We also show the Pareto front of graphs (of size $N=8$) that maintain diversity the longest for bd and db.
Further, we show that some graphs that quickly homogenize can maintain high levels of diversity longer than graphs that slowly homogenize.
For directed graphs, we give simple contracting star-like structures that have superexponential time scales for maintaining diversity.
\blfootnote{Corresponding author: dbrewster@g.harvard.edu}}

\section*{Significance Statement}
Evolution---either by genetic reproduction or by learning---occurs in populations.
The structure of a population affects the time scale and outcome of evolutionary processes.
The propensity of populations to maintain diversity is of great interest in evolutionary biology, ecology and social science.
Here we calculate for how long various population structures can maintain diversity under neutral evolution.
In this setting, diversity is lost by random drift. We give precise results for a large variety of structures.
We find that some structures have higher order polynomial or even superexponential timescales for maintaining diversity.
For realistic population sizes of thousands or millions of individuals, those structures can maintain diversity for times that exceed the lifetime of a universe.
Therefore, they protect diversity ``forever''.

\section*{Introduction}

Evolutionary graph theory is a method for studying the effect of population structure on evolutionary dynamics
\cite{
    Nowak_May_1992,
    nowak2003linear,
    lieberman2005evolutionary,
    nowak2006evolutionary,
    Ohtsuki_Hauert_Lieberman_Nowak_2006,
    tarnita2009evolutionary,
    allen2012evolutionary,
    allen2013spatial,
    diaz2021survey%
}.
The individuals occupy the vertices of graphs, and the edges specify interactions between individuals.
The special case of a well mixed population is given by a complete graph with identical weights.
In the case of constant selection, we are interested in the role of population structure on suppression/amplification selection effects and evolutionary timescales
\cite{
    adlam2015amplifiers,
    pavlogiannis2017amplification,
    pavlogiannis2018construction,
    tkadlec2020limits,
    allen2021fixation,
    tkadlec2021fast,
    abbara2024mutant,
    fruet2024spatial,
    svoboda2024amplifiers,
    kopfova2024colonization,
    kuo2024evolutionarytimes,
    kuo2024evolutionary%
}.
Weighted edges can create amplification or suppression effects
\cite{tkadlec2019population,tkadlec2021fast,bhaumik2024constant}. 
Isothermal graphs have the same fixation probability as the well mixed population
\cite{lieberman2005evolutionary,nowak2006evolutionary,adlam2014universality}.
Uni-directional edges can introduce exceedingly long absorption times
\cite{
    diaz2016absorption,
    brewster2024fixation%
}.
Even in neutral evolution, details of the evolutionary process can heavily affect timescales
\cite{IwamasaMasuda2014,mcavoy2018stationary,gao2024speed}.
Environments with mixed resource abundances have effects on fixation probabilities
\cite{kaveh2019environmental,kaveh2020moran}.
For frequency dependent selection, it is known that some graphs and update rules can promote evolution of cooperation
\cite{
    Nowak_May_1993,
    Nakamaru_Matsuda_Iwasa_1997,
    Hauert_Doebeli_2004,
    nowak2006five,
    ohtsuki2006evolutionary,
    taylor2007evolution,
    ohtsuki2008evolutionary,
    nathanson2009calculating,
    tarnita2009evolutionary,
    tarnita2009strategy,
    fu2010invasion,
    van2012direct,
    allen2014games,
    Allen_Lippner_Chen_Fotouhi_Momeni_Yau_Nowak_2017%
}.
Expected absorption times in structured populations have also been studied in continuous time
\cite{
    donnelly1983finite%
}.

In this paper, we analyze graphs for their ability to maintain diversity in neutral evolution.
We consider a population of finite size, $N$.
Initially all individuals represent different types.
All types have the same reproductive rate.
We ask: what is the expected time, $T_N$, until all individuals descend from the same type.
This time is known as total coalescence time in biology.
For a well-mixed population with $N$ individuals,
the coalescence time is known to be $N$ generations, or $N^2$ reproductive events 
\cite{
    nordborg2002separation,
    nordborg2019coalescent,
    allen2024coalescent%
}.
Our work builds upon and extends prior studies of absorption times in evolutionary dynamics.
Iwamasa \& Masuda \cite{IwamasaMasuda2014} study the consensus time of voter models on various graphs.
They show that for two types on small networks, the barbell and double star graph families maximize the expected absorption time for death-birth (db) and birth-death (bd) updating, respectively. Further, they calculate the asymptotic expected absorption times for two types on those graphs.
Diaz et al. \cite{diaz2016absorption} present a directed graph family that has at least exponential absorption time if one of the two types has a fitness advantage.
Gao et al. \cite{gao2024speed} analyze all undirected graphs of size $N=6$. They compute the absorption time starting with two types under neutral evolution and db updating. They find that graphs with a bottleneck between two large components typically have a large absorption time.

In contrast, our study expands the scope to $N$ types and provides rigorous analyses of the fastest and slowest graphs.
We also examine directed graphs, which can superexponentially broaden the diversity timescales.
We give upper and lower bounds for the diversity time of any undirected or directed graph.
In addition to our proofs for arbitrarily-sized population structures,
we analyze various evolutionary properties of graphs up to size $N=100$.
In particular, we are concerned not only with the expected time until homogeneity,
but also the expected number of types remaining in the population as a function of time.
We examine tradeoffs in diversity time between db and bd updating relative to the population structure.


\section*{Results}

\subsection*{Diversity in structured populations.} 

Consider a population of $N$ individuals. Initially each individual is of a different type. Every time step, one individual is selected for birth and one individual for death. The individual selected for death is removed from the population. The individual selected for birth creates a copy of itself at the location of the individual that was selected for death. After many steps, the population will become homogeneous, which means that all individuals are of the same type. Once the population is homogeneous it remains so. Thus, homogeneity is an absorbing state. In principle, there are $N$ different absorbing states---one for each of the types that are present initially. We are interested in calculating the average time $T_N$ until one of the absorbing states is reached. 

Population states with more than one type are called heterogeneous (or diverse). All heterogeneous states are transient. They will be lost after some time. The absorption time $T_N$ gives us a measure for the ability of a population structure to maintain diversity.

We are interested in exploring population structures for their ability to maintain diversity for extended periods of time. We describe the population structure as a strongly connected directed graph $G=(V,E)$.
The vertices $V$ denote the locations of individuals in the population.
The edges $E$ represent possible interactions between the individuals. For any two individuals, $u$ and $v$, if $v$ is a neighbor of $u$ in $G$, then the offspring of $u$ can replace $v$. The evolutionary dynamics on graphs can be interpreted as biological reproduction or learning. In the case of learning, one individual becomes a learner and the other a teacher. Then the learner adopts the type of the teacher.

The order of birth and death matters. Under birth-death (bd) updating, an individual $i$ is chosen uniformly at random from the population to reproduce.
Then an individual $j$ is chosen for death uniformly at random from the outgoing neighbors of $i$.
If individual $i$ resides at location $u\in V$ and $j$ resides at location $v\in V$,
then the probability of this event occurring at any step given $(u, v)\in E$ is
\begin{equation}
    \frac1N \cdot \frac{1}{\deg^+(u)}
\end{equation}
Here $\deg^+(u)$ represents the number of outgoing neighbors of vertex $u$.

Under death-birth (db) updating, an individual $j$ is chosen uniformly at random from the population to die.
Then an individual $i$ is chosen for birth uniformly at random from the incoming neighbors of $j$.
If individual $j$ resides at location $v\in V$ and $i$ resides at location $u\in V$,
then the probability of this event occurring at any step given $(u, v)\in E$ is
\begin{equation}
    \frac1N\cdot \frac{1}{\deg^-(v)}
\end{equation}
Here $\deg^-(v)$ represents the number of incoming neighbors of $v$.
\begin{figure}
   \centering 
   \includegraphics[width=1\textwidth]{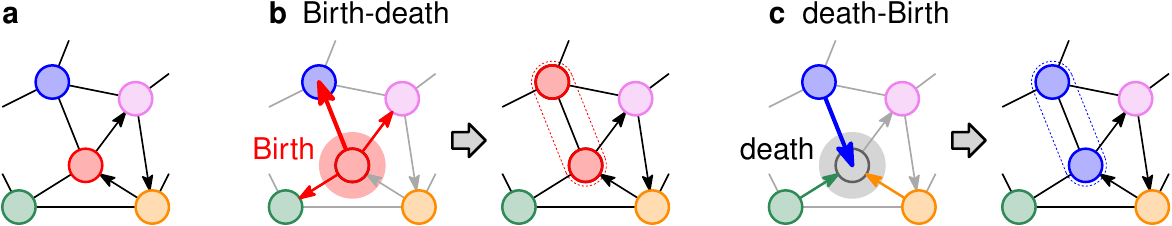}
   \caption{\textbf{a,} An example of a directed graph with various types in the population. %
   \textbf{b,} For birth-death (bd) updating, first an individual is chosen for reproduction and then one of its neighbors is chosen to be replaced; here the center (red) vertex is selected for birth and the blue vertex is selected for death. %
   \textbf{c,} For death-birth (db) updating, first an individual i chosen for death (or to update its type) and then one of the neighbors is chosen for reproduction; here the center (grey) vertex is selected for death and the blue vertex is selected for birth. %
   }
   \label{fig:bd-db-example}
\end{figure}

We call a graph undirected (bi-directional) if $(u, v)\in E$ implies $(v, u)\in E$ for all vertices $u,v\in V$.
In other words, individuals have reciprocal interactions in undirected graphs.
Since the number of incoming neighbors is the same as the number of outgoing neighbors in an undirected graph,
we denote the number of neighbors of a vertex $u\in V$ in an undirected graph as simply $\deg(u)$.

An undirected graph where all vertices have the same number of neighbors is called a regular graph.
Suppose all vertices in a regular graph have $D$ neighbors.
Then assuming $(u,v)\in E$,
the probability that location $u\in V$ is selected for birth and location $v\in V$ is selected for death is given by
\begin{equation}
    \frac{1}{N}\cdot \frac{1}{D}
\end{equation}
Since this relationship holds regardless of the update rule,
questions about diversity on regular graphs are unaffected by the governing dynamics.
See \Cref{fig:bd-db-example} for illustrations of the two update rules.

\subsection*{Time of evolution}
\begin{figure}
    \centering
    \includegraphics[width=\linewidth]{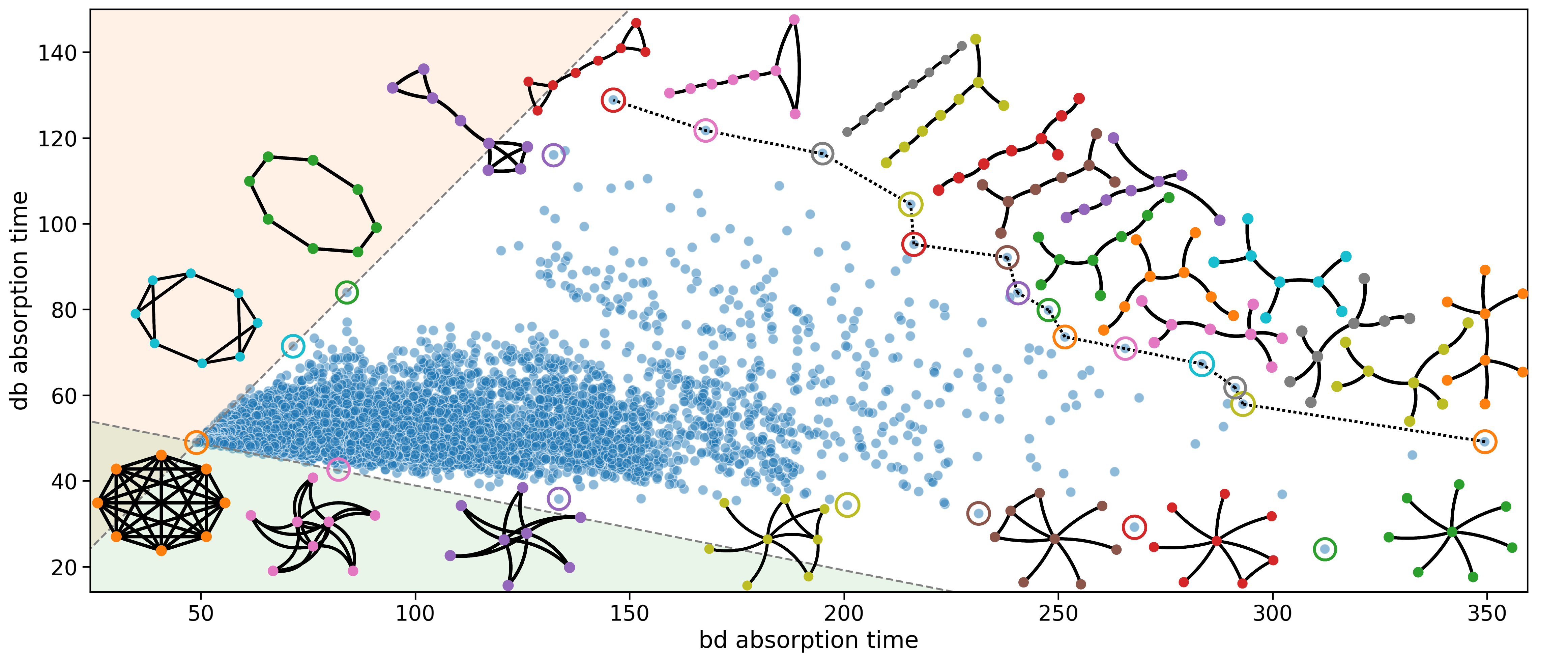}
    \caption{Absorption times for bd vs db updating for all $11117$ connected undirected graphs with $N=8$ vertices. %
    Each blue dot is a graph. %
    The positively sloped grey dashed line has unit slope and passes through the dot representing the complete graph; %
    all regular graphs are on this dashed line.
    The negatively sloped grey dashed line is such that it is
    the smallest sector that contains all the dots and has apex at the dot that corresponds to the complete graph.
    Some dots are circled with its corresponding graphical representation in the same color.
    The double star (orange) maximizes the bd absorption time. 
    The barbell (red) maximizes the db absorption time.
    The Pareto front (the dashed black lines) connects the two.
    }
    \label{fig:n8}
\end{figure}

\begin{figure}
    \centering
    \includegraphics[width=1\linewidth]{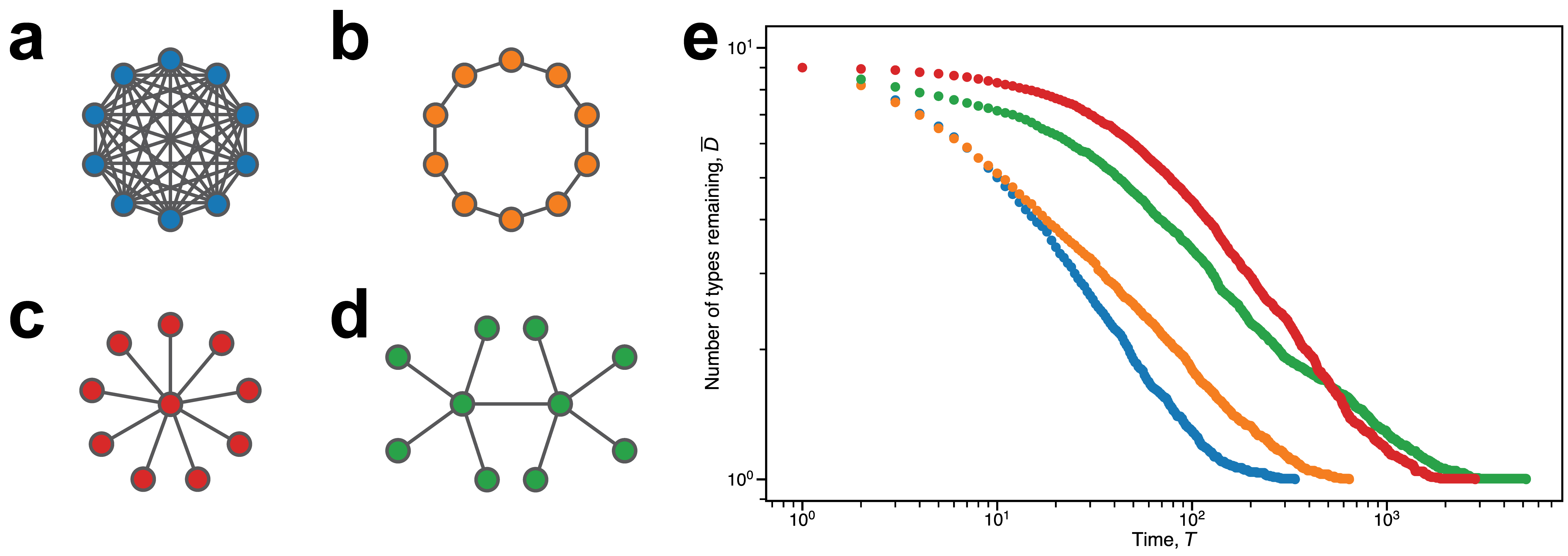}
    \caption{%
        Various undirected graph families on $N=10$ vertices: %
        \textbf{a}, the complete graph; %
        \textbf{b}, cycle; %
        \textbf{c}, star; and %
        \textbf{d}, double star. %
        \textbf{e},
        Results of average number of types remaining in the population, $\overline{D}$, at time $T$, averaged over $250$ simulations
        of birth-death updating
        per graph.
        Each graph has $N=10$ vertices.
        The plot is on a log-log scale to accentuate the number of types remaining when the values are close to one another.
        The horizontal axis begins at $T=1$.
        If $\overline{D}=1$ at a particular time $T$, no dot is drawn.
        Note that on average the star graph maintains more diversity than the double star graph up until roughly $T\approx 5\times 10^2$.
        See Figure S1 in the \SI{} \SIExtendedDataFigures{} for the plot when $N=100$.%
    }
    \label{fig:diversity-different-graphs-cc}
\end{figure}

We want to calculate the time (in number of steps) until the population becomes homogeneous.
Since the population starts with maximum diversity, we measure the ability of a population structure to maintain diversity by the expected time until homogeneity is reached.
We refer to the time until homogeneity as the absorption time.
It is important to note that the absorption time of an observed bd or db process is a number
while the absorption time of a population structure is a random variable.
We are interested in the expected absorption time of various population structures
(see \Cref{fig:n8}, \Cref{fig:diversity-different-graphs-cc}, and \Cref{table} for a few examples).
We measure time as a function of the population size $N$.
Often, we are interested in the asymptotics of the time rather than an exact expression.


\subsection*{Well-mixed populations.}

Consider a population structure with $N$ individuals where every individual interacts with every other individual.
This configuration is assumed when analyzing evolutionary dynamics without explicit reference to population structure.
We represent this well-mixed population structure as a complete graph on $N$ vertices with self-loops:
the edges are the Cartesian product of the vertex set by itself.
Complete graphs are highly symmetric.
Thus knowledge about the frequencies of the various types in the populations
is sufficient information for calculating the expected absorption time.

\begin{table}[h]
    \centering
    \begin{tabular}{c l c c||} 
        \hline
        & Graph family & bd time & db time \\ [0.5ex] 
        \hline\hline\hline
        \multirow{6}{*}{\rotatebox{90}{\parbox{2cm}{\centering \textit{\tiny Undirected}}}} 
        & Complete & $\Theta(N^2)$ & $\Theta(N^2)$ \\
        & Cycle & $\Theta(N^3)$ & $\Theta(N^3)$ \\
        & Star & $\Theta(N^3)$ & $\Theta(N\log N)$ \\
        & Double star & $\Theta(N^4)$ & -- \\
        & Barbell & -- & $\Omega(N^4)$ \\
        & Undirected graph & $O(N^6\log N)$, $\Omega(N\log N)$ & $O(N^5\log N)$, $\Omega(N\log N)$ \\ [0.5ex]
        \hline\hline
        \multirow{3}{*}{\rotatebox{90}{\parbox{0cm}{\centering \textit{\tiny Directed}}}} 
        & Contracting star & $2^{\Theta(N\log N)}$ & $2^{\Theta(N\log N)}$ \\
        & Directed graph & $2^{O(N\log N)}$ & $2^{O(N\log N)}$, $\Omega(N\log N)$ \\
        \hline
    \end{tabular}
    \caption{%
        Asymptotic absorption times for various graph families under bd and db updating.
        We use $O$, $\Omega$, and $\Theta$ to represent asymptotic upper, lower, and tight bounds, respectively
        (see \S 1.2 of \cite{arora2009computational} for formal definitions).
        For the double star, we have no estimate for db updating.
        For the barbell, we have no estimate for bd updating.
        For bd updating, there is a gap between the slowest undirected family of graphs we know (double star)
        and our theoretical upper bound for any undirected graph.
        Similarly, for db updating there is a gap between the slowest undirected family of graphs we know (barbell)
        and our theoretical upper bound for any undirected graph.
        In both cases, whether our analysis is not tight enough or there are even slower graph families that we have not found is unknown.
        In contrast, we find that contracting stars are the slowest of the directed graphs.
    }
    \label{table}
\end{table}

Consider birth-death updating and suppose there are currently $k$ types in the population.
Let $\lambda\equiv(\lambda_1,\ldots,\lambda_k)$ denote the vector of abundances.
We have $\lambda_1 + \cdots + \lambda_k = N$.
We order the abundances such that $\lambda_1 \geq \lambda_2 \geq \cdots \geq \lambda_k > 0$.

During each time step, one of three events occurs:
\begin{enumerate}
    \item the abundances, $\lambda_1,\ldots,\lambda_k$, remain exactly the same;
    \item the abundance of one type increases by one, while the abundance of another type decreases by one,  but the number of types in the population remains the same;
    \item the abundance of one type increases by one, the abundance of another type decreases by one, and the number of types in the population decreases by one.
\end{enumerate}
The process can be described as beginning in a state of maximum entropy ($\lambda_1=\cdots=\lambda_N$)
and reaching a state of minimum entropy ($\lambda_1=N$).
We show that the expected absorption time of this process starting from configuration $\lambda$ is exactly 
\begin{equation}\label{eq:well-mixed-formula}
    T_N= N^2 - N - \sum_{i=1}^k\sum_{\ell=1}^{\lambda_i-1}\frac{(N+\lambda_i-2\ell)\ell}{N-\ell}
\end{equation}
One intuitive way to think about the formula is as follows (see \SI{} \SIComplete{}).
Imagine the ``histogram'' of the partition $\lambda$ and for each $h\ge 0$,
denote by $b_h$ the number of boxes above the line $y=h$. (In particular, $b_0=N$.)
An explicit formula for $b_h$ is $\sum_{i} \max(\lambda_i-h,0)$.
Then the expected absorption time from the given configuration is
\begin{equation}\label{eq:histogram}
    T_N=N\cdot\left(N - \sum_{h=0}^{N-1} \frac{b_h}{N-h}\right)
\end{equation}
In the case when $k=N$ (i.e. $\lambda_1=\cdots=\lambda_N$),
Equation~\eqref{eq:well-mixed-formula} yields the expected absorption time from maximum diversity as
\begin{equation}
   N\cdot(N-1)
\end{equation}
If we consider a complete graph with no self-loops, the expected absorption time is
$(N-1)^2$ which is not asymptotically different than the expected absorption time with self-loops
(See the \SI{} \SIComplete{} for details).
Well-mixed population structures are represented by regular graphs since each vertex has the same number of neighbors.
Thus our results under birth-death dynamics are the same as the results for death-birth updating.
Next, we will explore population structures beyond well-mixed populations.

\subsection*{Cycles.}
Consider a population with $N$ individuals whose locations form a circle-like structure.
Individuals interact with each of their two adjacent neighbors.
This population structure is represented by an undirected cycle graph.
A cycle is a regular graph since each vertex has exactly two neighbors.
Types are always clustered together on the vertices of the cycle.
As individuals give birth and die, some clusters take over others.
In many steps of the selection process on the cycle,
individuals in the interior of the cluster are chosen for reproduction;
the individual is only able to reproduce to locations where individuals of its type already reside.
Thus no change in the population configuration occurs.
When the individual selected for birth resides on a boundary between differing types,
there is a $50\%$ chance that the configuration of the population changes.
These updates are called active steps.

Individuals of the same type are always clustered together on the cycle.
Using similar logic to the case of the well-mixed population, it suffices to know only the frequencies of 
the types and their relative locations around the perimeter of the cycle.
Recall $b_h = \sum_{i} \max(\lambda_i-h,0)$.
For the expected absorption time from given frequencies, we arrive at 
\begin{equation}
    T_N=\frac{(N+1)N(N-1)}6-\sum_{h=0}^{N-1} b_h\cdot h
\end{equation}
When the process starts with maximum diversity, this results in the expected absorption time as simply
\begin{equation}
    T_N=\frac{(N+1)N(N-1)}6
\end{equation}
Some existing results for cycles are known~\cite{broom2010evolutionary}.
See the \SI{} \SICycle{} for more details.

So far we have examined regular graphs.
Next we will analyze graphs that are far from regular.

\subsection*{Stars.}
A star graph has one central vertex and multiple vertices connected to the central vertex.
For a star with $N$ vertices we denote $n$ as the number of non-central vertices so that $n+1 = N$.
More formally, the central vertex $c\in V$ is connected to the remaining vertices $v_1,\ldots,v_n\in V$ bi-directionally.
These graphs are not regular because the central vertex has degree $n$ whereas the remaining vertices have degree $1$.
However, stars are very similar to complete graphs in the following sense:
the non-central vertices are connected to each other with paths of length two.
There are two types of events that can occur on a star:
\begin{enumerate}
    \item\label{star:periphery} a vertex on the periphery is selected for birth, or
    \item\label{star:central} the central vertex is selected for birth.
\end{enumerate}
First, consider birth-death updating on a star.
The case of Event~\ref{star:periphery} occurs with probability $1-1/N$.
The only place for a vertex on the periphery to give birth is into the center.
On the other hand, the case of Event~\ref{star:central} happens with probability only $1/N$.
When the center is selected for reproduction, it places its offspring at a vertex chosen uniformly at random from the periphery.
On average, the center reproduces every $N$ steps.
The type of the individual at the central vertex at the time it gives birth is highly likely to be directly proportional
to the relative abundances of the types on the periphery.
Thus the process is akin to the complete graph with each step scaled by a factor of $N$.
We show that the expected absorption time of a star on $N$ vertices is $\Theta(N^3)$.
We conjecture that the exact expected absorption time under birth-death updating is
\begin{equation}
    T_N = n^3 - n^2 + n\cdot H_n
\end{equation}
The expression $H_n$ represents the sum of the first $n$ terms of the harmonic series $\frac11 + \frac12 + \frac13 + \cdots$.
See the \SI{} \SIStar{} for more details.

Next, consider death-birth updating on a star.
The case of Event~\ref{star:periphery} only happens when the central vertex is selected for death.
This event occurs with probability $1/N$.
On the other hand, Event~\ref{star:central} occurs with remaining probability.
Let $i$ be the number of types in the periphery different from the center.
There are two kinds of active steps. 
Either the center is replaced by a different type
or the center reproduces onto a different type.
The former event has probability $(1/N)\cdot (i/n)$
and the latter event has probability $i/N$.
The ratio between the two events is $1/n$.
Thus in $n$ active steps, the center is not replaced by a different type with probability 
\begin{equation}
    \left(1-\frac{1}{N}\right)^n \ge 1/e
\end{equation}
If the center is replaced, we restart the process.
That means with constant probability, the process ends in $n$ active steps.
Counting all steps gives a logarithmic slowdown yielding an expected absorption time of $\Theta (N \log N)$ under death-birth dynamics.
See the \SI{} \SIStar{} for details.

The star is the first population structure we have seen that is sensitive to the dynamics.
Stars promote diversity under birth-death updating and demote diversity under death-birth updating.

\subsection*{Double stars.}
A double star graph is a bi-directional graph composed by joining two equally sized stars together by their central vertices.
For simplicity, we consider only double stars with an even number of vertices.
We denote $n$ as the number of non-central nodes on one star.
The total number of nodes in a double star is $N=2n+2$.
Similar to stars, double stars are also non-regular graphs since the two central vertices have degree $n+1=N/2$ and the remaining nodes have degree $1$.
We consider birth-death updating.
Compared to the star, there is an additional event that could happen:
one center could give birth onto the other center.
This invasion attempt happens with probability roughly
\begin{equation}
    2\cdot\frac{1}{N}\cdot\frac{1}{N/2} = \Theta\left(N^{-2}\right)
\end{equation}
Suppose a star has a homogeneous population except for a single differing type at its central vertex.
It is known that under birth-death updating, the probability the individual intitially placed at the center will take
over the population is
\begin{equation}
\frac{\frac1{n}}{\frac1{n} + n} = \frac{1}{1+n^2} = \Theta\left(N^{-2}\right)
\end{equation}
This probability is known as the fixation probability (see \cite{nowak2006evolutionary}).
Successful invasion is rare.
If the individual initially placed at the center goes extinct, this likely happens quickly, in a constant number of steps.
Since both an invasion attempt and fixation must occur, a successful invasion takes roughly $\Theta(N^{4})$ steps on average.
Thus the typical evolution of the population on a double star proceeds as follows:
\begin{enumerate}
    \item Evolution occurs primarily in the two stars of the double star; sometimes invasion attempts occur but invaders are quickly wiped out.
    \item After roughly $\Theta(N^3)$ steps the stars are each homogeneous, but there are still two types remaining in the population.
    \item The process terminates after the next successful invasion; this takes on average $\Theta(N^4)$ steps.
\end{enumerate}
Overall, the expected absorption time for a double star is $\Theta(N^4)$.
This structure promotes diversity under birth-death updating for the longest out of all of the undirected graphs we considered.
In db updating, the central hubs give birth often.
Though unlike the star,
the two stars on the double star must eventually agree on the type.
See the \SI{} \SIDoubleStar{} for proofs of the upper and lower bounds.

\subsection*{Barbells.}
\begin{figure}
    \centering
    \includegraphics[width=1\linewidth]{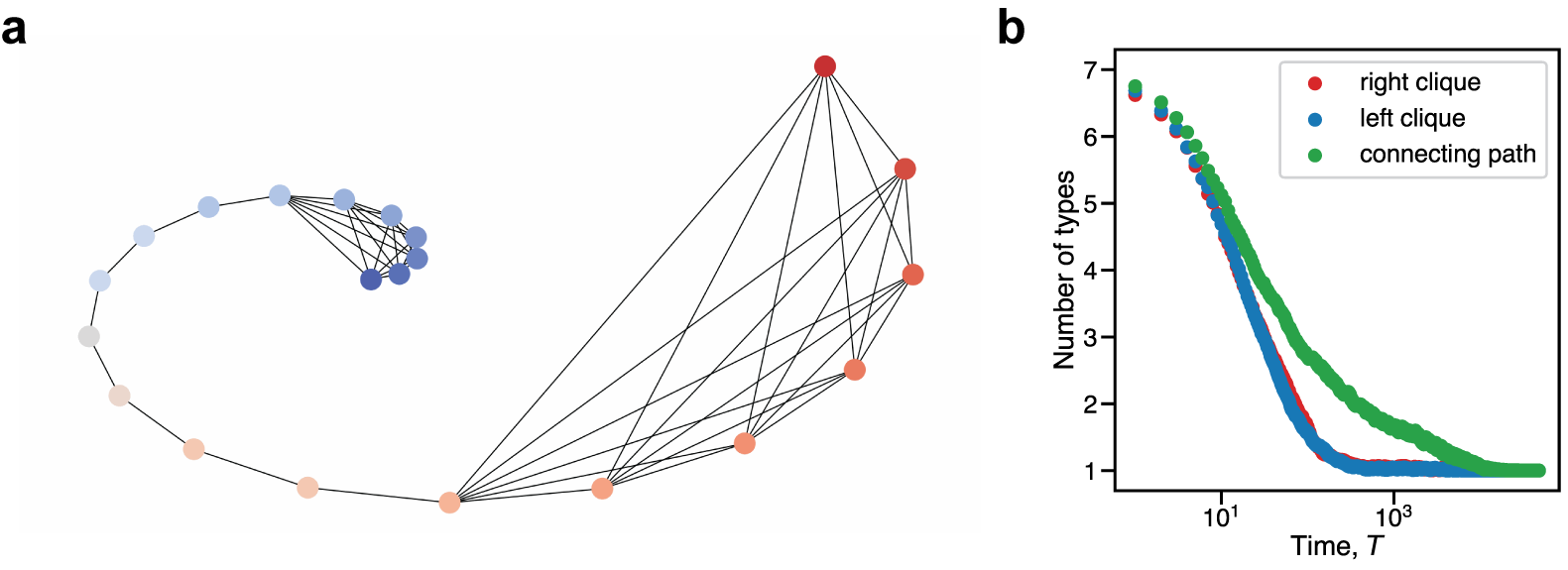}
    \caption{
        \textbf{a,} A barbell graph with $N=7+7+7=21$ vertices.
        \textbf{b,} A semi-log plot displaying the number of types in each part of the barbell graph versus time, averaged over $250$
        simulations.
    }
    \label{fig:barbell}
\end{figure}

A barbell graph is a bi-directional graph composed by joining two equally sized cliques
(i.e., fully connected subgraphs)
together by a path.
If the two cliques have $n$ vertices,
then the path has $n$ vertices,
combining for a total of $N=3n$ vertices in the graph (see \Cref{fig:barbell}).
For death-birth updating, the process eventually settles on two types in the population.
Each type occupies a clique and part of the path.
Then one type attempts to invade the other clique.
There is a $1/n$ chance invasion is successful.
But invasions only happen roughly every $n^3$ steps due to the absorption time of a path graph.
The ends of the paths have high degree. 
Thus it is much more likely that the path remains heterogeneous when it is nearly homogeneous due to the death-birth 
updating process.
The process resolves in expected time $\Omega(N^4)$.
See the \SI{} \SIBarbell{} for details.
For birth-death updating,
the barbell diversity time is faster.
Invasion into a clique happens at a much higher rate for birth-death updating since
an end node on the connecting path has a $1/2$ probability of invading if it selected for birth (with probability $1/N$).
However for death-birth updating, invasion occurs when a node in a clique connected to the connecting path dies (with probability $1/N$);
but, there is only a $1/n$ probability that the node on the path will invade into the clique.

\subsection*{Time bounds on any two-way population structure.}
Under birth-death updating, it is known that if the initial configuration on an undirected graph contains only two types,
the expected absorption time is $O(N^6)$.
Recent work for a multi-type birth-death process yields an $O(N^7)$ upper bound for the expected absorption time when
the process starts with $N$ types \cite{goldberg2024parameterised}.
We show that this upper bound can be tightened to $O(N^6\log N)$ by a divide-and-conquer proof strategy.
For death-birth updating, the literature on consensus problems gives an $O(N^5)$ upper bound on the
expected absorption time when the process starts with two types~\cite{cooper2016linear}.
Similar to the birth-death case, we can achieve an upper bound on the expected absorption time for any graph of $O(N^5\log N)$ for death-birth updating.
See the \SI{} \SIGeneral{} for details.

For both birth-death and death-birth updating, an $\Omega(N\log N)$ lower bound for the expected absorption time follows by considering that at least $N-1$ locations
must eventually be a death site for the process to absorb.
The expected amount of time for $N-1$ locations to be a death site is $(N-1)\cdot H_{N-1}$.
We note that $H_{N-1} = \Theta(\log N)$.
See Methods and \SI{} \SIGeneral{} for more details and proof.


\subsection*{Contracting star}
We have shown that all undirected graphs have expected absorption time at most some polynomial function of the population
size.
We call these absorption times short.
In neutral evolution under birth-death updating, it is known that some families of directed graphs (graphs that have some one-way connections)
also have short absorption times \cite{brewster2024fixation}.
We give a construction of a directed graph family with long absorption times, times that are some superexponential function of the population size.

\begin{figure}
    \centering
    \includegraphics[width=1\linewidth]{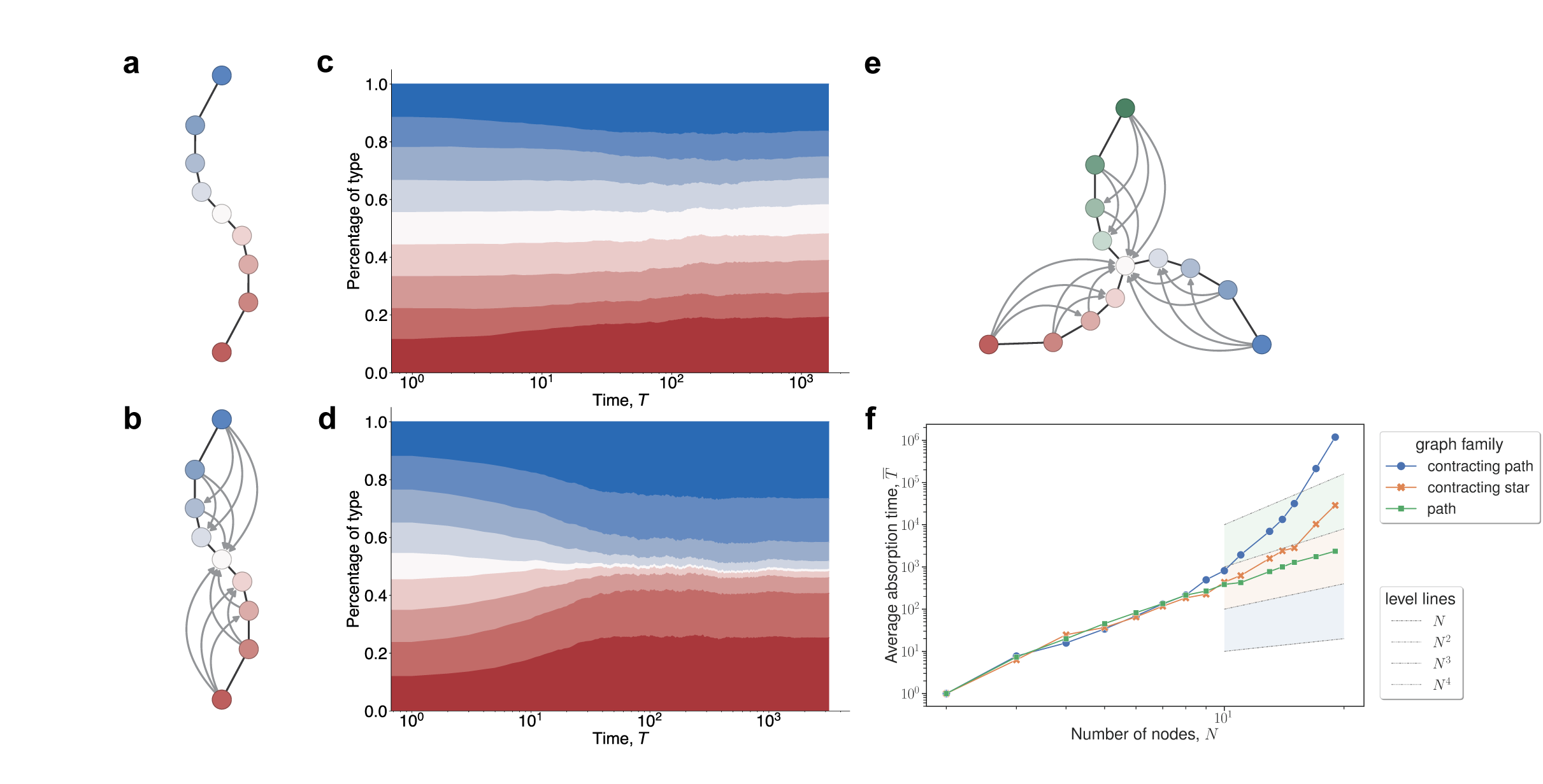}
    \caption{
        \small{
        \textbf{a,} An undirected path on $N=9$ vertices.
        \textbf{b,} A contracting star (or contracting path) on $N=9$ vertices and $b=2$ blades.
        A contracting path is a composition of a bidirectional (undirected) path and
        ``contracting'' directed edges pointing inwards.
        \textbf{c,} Plots for the normalized frequency of a type in the population of the undirected path on $N=9$ vertices versus time over $1000$ simulations.
        The population starts with the types as colored in (\textbf{a}).
        The height of each color is proportional to frequency of the corresponding type in the population.
        \textbf{d,} Plots as in (\textbf{c}) but for the contracting path on $N=9$ vertices.
        \textbf{e,} Contracting star on $N=13$ vertices and $b=3$ blades.
        \textbf{f,} Simulation results of the expected absorption time for birth-death updating of a undirected path, a contracting path, and a contracting star (with three blades)
        over varying population sizes.
        The plot is semi-log and $N$ ranges from $2$ to $20$.
        Each dot represents the average of $100$ trials starting from maximum diversity.
        The dotted lines towards the right side of the plot indicate various power law level lines.
        The path graph family follows a level line, but the contracted star graph families grow faster than
        some polynomial of the population size.
        }
    }
    \label{fig:contracting-star}
\end{figure}

A contracting star is a directed graph with multiple blades bi-directionally connected to a central vertex.
Each blade consists of a bidirectional path.
For every pair of vertices on a blade, there is a directed edge from the vertex farther from the center to the vertex closer to the center
(see \Cref{fig:contracting-star}).
For simplicity, we restrict contracting stars to have $N$ vertices and $b$ equally sized blades
so that $b$ divides $N-1$.
We show that under birth-death updating, the expected absorption time of a contracting star with two blades is
\begin{equation}
    T_N \geq 2^{\Omega(N \log N)}
\end{equation}
See the \SI{} \SIDirected{} for more details.

\subsection*{Computer Experiments}
\begin{figure}
    \centering
    \includegraphics[width=\linewidth]{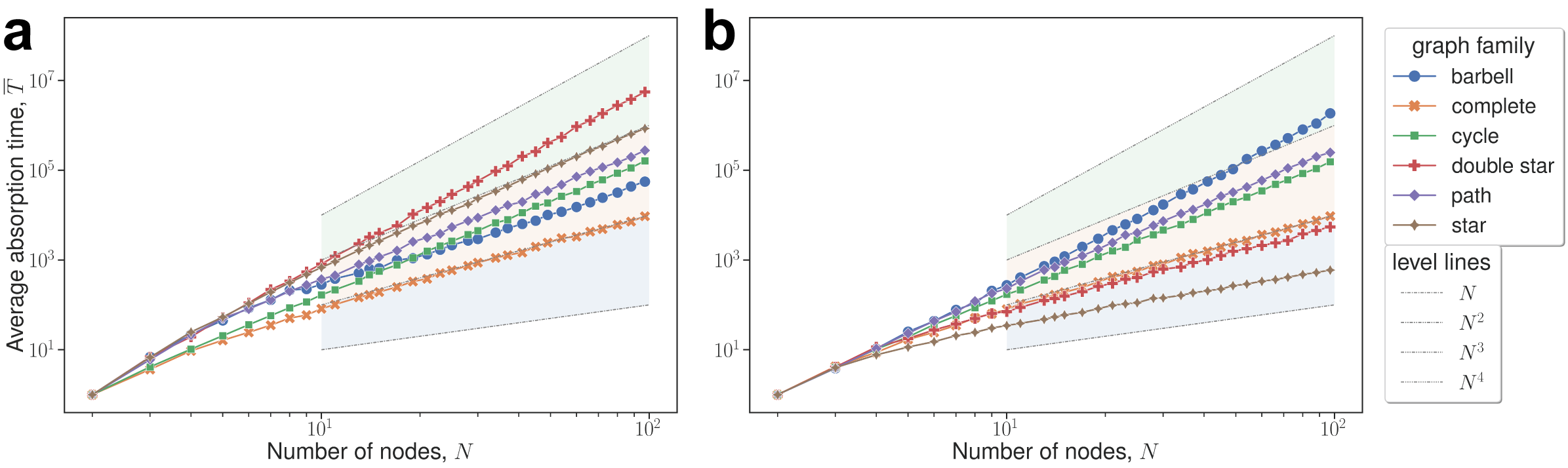}
    \caption{%
    Log-log plots of the absorption time ($T$) of various graph families for $1 \leq N \leq 100$, with $250$ simulations per data point.
    The boundaries of the shaded regions denote the power law exponent; these level lines are described in the legend on the bottom right.
    We can see that regular graphs (e.g., complete and cycle graphs) are unaffected by the updating dynamics.
    \textbf{a,} birth-death updating
    \textbf{b,} death-birth updating.
    }
    \label{fig:trends}
\end{figure}

We investigate properties of small graphs.
First, we consider various graph families with $N\leq 100$ vertices
and estimate the expected absorption times via simulations.
For bd updating (see \Cref{fig:trends}\textbf{a}), we see that double stars are asymptotically
slower than stars, which is in alignment with our theoretical results.
Similarly, stars seems asymptotically slower than cycles and paths.
Finally, cycles and paths seem asymptotically slower than well-mixed populations.
It appears that paths are slower than cycles by some multiplicative constant.
In contrast, for db updating (see \Cref{fig:trends}\textbf{b}), we see that double stars are not the asymptotically slowest
graphs presented.
We do not rigorously analyze double stars under death-birth updating,
but from \Cref{fig:trends}\textbf{b} we see that the diversity time of the double star
becomes much faster under death-birth updating compared to birth-death updating.
We also see that stars absorb faster than complete graphs.
It is known that the absorption times for regular graphs (e.g., complete graph, cycles) are independent of the
two updating mechanisms we consider.

\begin{figure}
    \centering
    \includegraphics[width=1\linewidth]{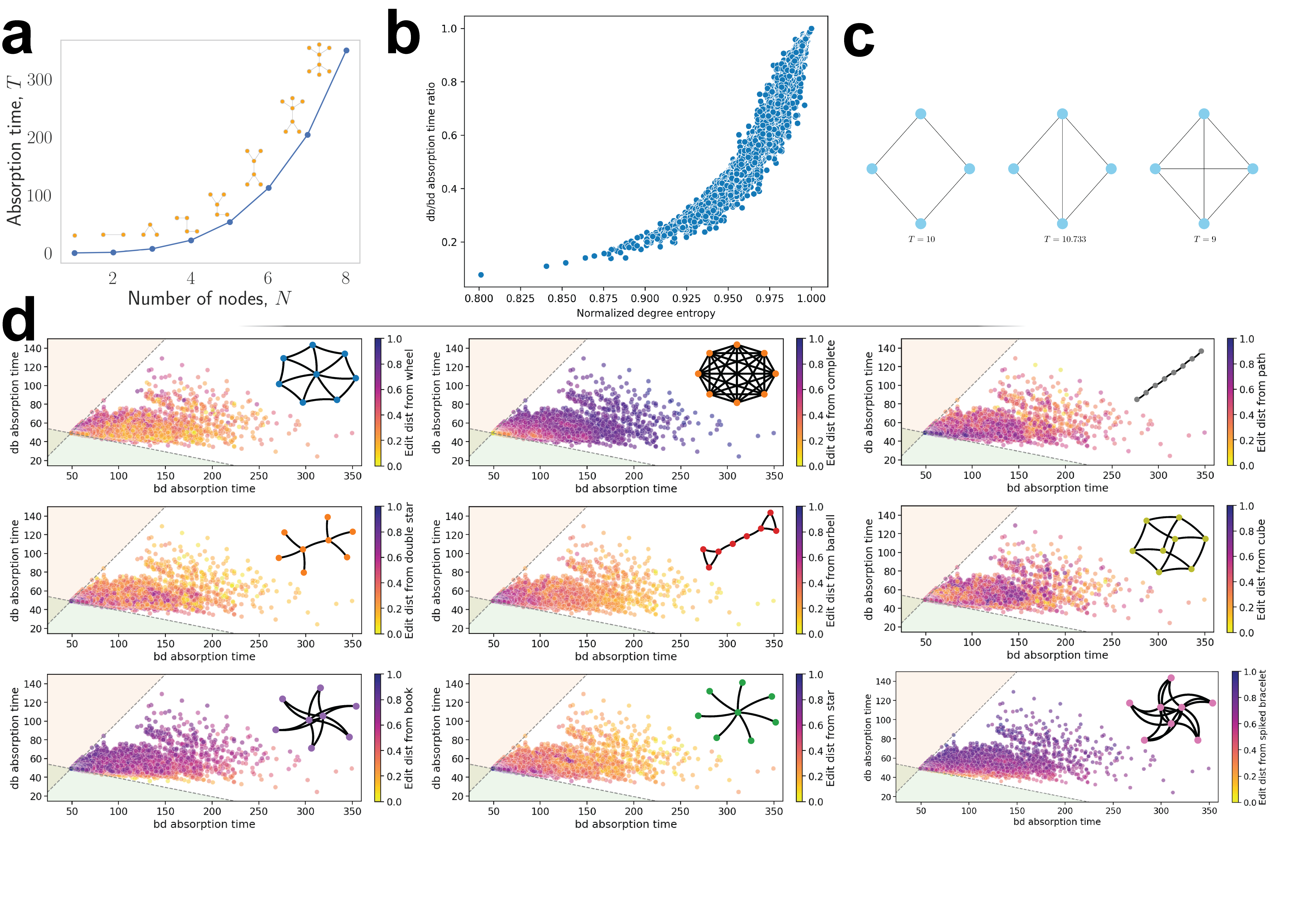}
    \caption{
    \textbf{a,} Absorption times ($T$) of the slowest undirected graphs of birth-death updating for $1 \leq N \leq 8$.
    \textbf{b,} db absorption time divided by bd absorption time (db/bd absorption time ratio),
    versus normalized degree entropy for all graphs with $N=8$ vertices. 
    \textbf{c,} Three graphs on $N=4$ vertices with corresponding bd absorption times below.
    We see that adding an edge can increase the absorption time (compare the leftmost graph absorption time to that of the center graph).
    We also see that adding an edge can decrease the absorption time (compare the center graph absorption time to that of the rightmost graph).
    \textbf{d,} Absorption times under bd and db updating for $N=8$.
    The colors of the dots indicate the graph edit distance from a target graph.
    Lighter colors indicate a closer distance to the target whereas darker colors indicate a farther distance.
    }
    \label{fig:slowest}
\end{figure}

Next, we look at all connected undirected graphs with $N=8$ vertices.
We analyze the bd absorption time versus the db absorption time for such graphs.
From \Cref{fig:n8} we observe that there are no graphs that have a higher db absorption time than bd absorption time;
the graphs that come closest are the regular graphs for which the two times are exactly equal.
The edit distance between two graphs is the minimum number of vertex (or edge) deletions (or insertions) to transform
one graph into an isomorphic version of the other.
The Pareto front of the graphs with the longest absorptions under bd or db seems to fall under a gradient from barbell
to double star, with small graph edit distance between consecutive members.
We also see that distance in the bd vs db absorption time plane is not always correlated with the graph edit distance
(see \Cref{fig:slowest}\textbf{d}).
The normalized degree entropy of an undirected graph is defined as
\begin{equation}
    -\frac{1}{\log N}\sum_{u\in V} \frac{\deg(u)}{\sum_{v\in V} \deg(v)} \log\left(\frac{\deg(u)}{\sum_{v\in V} \deg(v)}\right)
\end{equation}
The normalized degree entropy measures the regularity of the graph;
the value of this quantity is between $0$ and $1$.
In \Cref{fig:slowest}\textbf{b}, we see that low normalized degree entropy
roughly correlates to lower db absorption times versus bd absorption times.
For $N=8$, the graph with the lowest normalized degree entropy is the star.
Graphs with higher normalized degree entropy are more time-robust to the particulars of 
the two updating mechanisms.
We note that adding edges to a graph does not necessarily decrease its expected absorption time
(see \Cref{fig:slowest}\textbf{c}).

For all connected undirected graphs with $N\leq 8$,
we computed the exact expected absorption time under bd updating.
We found that the slowest absorbers are graphs resembling double stars (see \Cref{fig:slowest}\textbf{a}).
This may indicate that double stars are the slowest absorbers under bd updating.

Finally, we more closely examine bd updating on various graphs of the same size.
We are interested in the average number of type remaining at time $T$ in the process.
We find that although a double star has a longer expected absorption time,
a star can maintain more types for a longer amount of time (see \Cref{fig:diversity-different-graphs-cc}).

\section*{Discussion}

In summary, we have shown that population structure and update rules can have large effects on the maintenance of diversity.

We have shown the following five results which hold both for birth-death (bd) and for death-birth (db) updating:
(i) For the complete graph, which describes a well mixed population, the time scale for loss of diversity is  $\Theta(N^2)$.
(ii) For the cycle, which describes a simple one-dimensional population structure, the time scale is $\Theta(N^3)$. 
(iii) The lower provable bound for any undirected graph is $\Omega(N \log N)$; for db updating the star matches this lower bound but we have no matching example
for bd updating.
(iv) The slowest directed graph which we have identified so far---the contracting star---loses diversity at the vast time scale of $2^{\Theta(N \log N)}$.
(v) The upper bound for the directed graph is $2^{O(N \log N)}$.

We derive the following additional results which hold for bd updating:
(i) the star has time scale $\Theta(N^3)$;
(ii) the double star has time scale $\Theta(N^4)$;
(iii) for any undirected graph the upper bound is $O(N^6 \log N)$.

We derive the following additional results which hold for db updating:
(i) the star has time scale $\Theta(N \log N)$;
(ii) the barbell has a lower bound of $\Omega(N^4)$;
(iii) for any undirected graph the upper bound is $O(N^5 \log N)$.

For bd updating, we establish that double-star graphs have a time scale of $\Theta(N^4)$,while any undirected graph has an upper bound of $O(N^6 \log N)$.
Closing this gap remains an open challenge.
Furthermore, the complete graphs has a time scale of $\Theta(N^2)$ while the lower bound for any graph is $\Omega(N \log N)$.
Determining whether tighter bounds can be achieved for intermediate cases is an intriguing direction for future research.
While we prove that star graphs have diversity times of $\Theta(N^3)$, we conjecture an exact formula for their diversity time.
Proving this formula would provide a deeper understanding of star dynamics.

For db updating, we show that barbell graphs have a lower bound of $\Omega(N^4)$ and that any graph has an upper bound of $O(N^5 \log N)$. Closing this gap remains an open challenge.
Star graphs have an upper bound of $O(N \log N)$ matching the general lower bound of $\Omega(N \log N)$ for any graph.

The superexponential time scale that is achieved by contracting stars means that diversity can by maintained ``forever" if the population size is not too small. But even the $N^4$ time scale that is reached by undirected graphs, which corresponds to $N^3$ generations, would imply that for (microbial) population sizes of $N=10^6$ diversity is maintained for $10^{18}$ generations, which exceeds the time scale of evolution on earth.

Our study suggests many possibilities for future research.
We have analyzed the expected time until a population becomes homogeneous.
However, we noticed that a graph could maintain some number of types longer than another but still homogenize quicker.
For example, we observed that a double star becomes homogeneous slower than a star on average,
but a double star loses types more rapidly than a star at times closer to the inception of the process.  
It would be insightful to understand which population structures can maintain diversity of at least a certain number of distinct types for the longest.

Also, the notion of diversity which we have used here is only one possibility of many.
We have counted the number of distinct types that is present in the population.
Other notions of diversity---such as the Simpson index or the Shannon entropy---take into account the frequency of different types and/or the spatial clustering of types. 

We plan to study the effect of mutation on maintaining diversity. In this setting new types are produced by mutation and existing types become extinct by random drift. Then the population reaches a steady state level of diversity. We ask: how does population structure affect diversity at steady state. 

Finally, one should investigate how variation in fitnesses affects our results.
Our work in this paper solely examines neutral evolution.

These research directions collectively point toward a more comprehensive understanding of how population structure shapes evolutionary timescales.
The interplay between spatial organization, mutation, selection, and various metrics of diversity represents a rich space for theoretical exploration with significant practical implications for ecology, virology, cultural evolution, and other fields.

{\small 

\section*{Methods}

\noindent%
Next, we formulate our model and mathematical methods.
See the \SI{} \SIModel{} for further details and proofs.

\noindent\textbf{Model.}
We consider a population of $N$ individuals undergoing a selection process with drift.
The population structure is represented by an unweighted graph of $N$ vertices (nodes).
Individuals in the population reside on the vertices of the graph.
The edges between individuals can be either bi-directional (two-way) or uni-directional (one-way);
if all edges are bi-directional, we refer to the graph representing the population structure as undirected.
Initially, each individual is a unique type.
At each step of the evolutionary process, an individual is selected for birth
and an individual is selected for death.
The individual selected for death is removed from the population.
The individual selected for birth places a copy of itself at the location
of the individual selected for death.
Each state of the process can be represented by a vector $\mathbf x \equiv (x_1,\ldots,x_N)$,
where $x_i\in\{1,\ldots,N\}$ indicates the type residing at location $i$ on the graph.

\noindent\textbf{Dynamics details.}
We consider two different dynamics on the population.
\begin{itemize}
    \item \textbf{birth-death updating:}
    Under birth-death (bd) dynamics, individual $i$ is chosen for birth uniformly at random from the population to give birth.
    Then individual $j$ is chosen for death uniformly at random from the outgoing neighbors of $i$.
    \item \textbf{death-birth updating:}
    Under death-birth (db) dynamics, individual $j$ is chosen for death uniformly at random from the population to give birth.
    Then individual $i$ is chosen for birth uniformly at random from the incoming neighbors of $j$.
\end{itemize}
We note that birth-death and death-birth are typically stylized as Birth-death (Bd) and death-Birth (dB) in the literature of evolutionary dynamics \cite{svoboda2024amplifiers,tkadlec2020limits}. 
The capitalized letter in ``Birth'' (``Death'') signifies that the individual selected for birth (death) is chosen proportional to its fitness,
whereas the individual selected for death (birth) is chosen uniformly at random.
Our evolutionary dynamics model corresponds to neutral evolution and thus all individuals have the same fitness.
Therefore we do not capitalize any letters in the names of the updating rules.

\noindent\textbf{Absorption time.}
The process always reaches a homogeneous state where there is only one type in the population.
From a homogeneous state, no more state changes can occur.
These homogeneous states are the only absorbing states in the Markov chain describing the process.
The absorption time is the number of steps until an absorbing state is first reached.
The expected absorption times starting in state $\mathbf x$, denoted $\tau_{\mathbf x}$, are the solution to the system of linear equations
\begin{equation}\label{eq:abs-time-system}
    \tau_{\mathbf x} =
    \begin{cases}
        0&\text{if }x_1 = \cdots = x_N, \\
        1 + \sum_{\mathbf x'} p_{\mathbf x\rightarrow \mathbf x'} \cdot \tau_{\mathbf x'} &\text{otherwise}
    \end{cases}
\end{equation}
The expression $p_{\mathbf x\rightarrow \mathbf x'}$ is the probability of transitioning
to state $\mathbf x'$ in the next step given the current state is $\mathbf x$.
We are interested in the expected absorption time for $\mathbf x = (1,\ldots,N)$.
In general, Equation~\eqref{eq:abs-time-system} has exponential size and becomes 
intractable to solve for large $N$.

\noindent\textbf{Fixation probability.}
Fixation of type $i$ occurs when all individuals in the population are of type $i$.
We note that since the process has no mutation, a type that has taken over the population will remain fixated indefinitely.
The fixation probability of type $i$ is the probability that type $i$ fixates.
Fixation probabilities depend on the initial configuration of types and the governing dynamics.

\noindent\textbf{Graph families.}
We examine properties of various graph families.
Each graph in a graph family is indexed by its size $N$.
Well-mixed populations are represented by a complete graph with self-loops.
The cycle graph family consists of undirected graphs where each vertex is connected to exactly two other vertices, forming a single closed loop.
The star graph family consists of undirected graphs with a central vertex that is directly connected to all other vertices;
these peripheral vertices have no connections to each other.
The double star graph family consists of graph formed by joining two stars graph of the same size by connected their centers;
if $N$ is odd, the star sizes differ by one.
An undirected graph is regular if each vertex has the same number of neighbors.
Complete and cycle graph families consists of regular graphs whereas star and double star graph families
contain graphs that are not regular.
A contracting star is a directed graph consisting of a central vertex connected to multiple blades of the same size.
Each blade is made by taking a bidirectional path and, for each vertex, adding directed edges to vertices on its left.
Then the left most vertex of the blade is connected to the center vertex bidirectionally.

\noindent\textbf{Mass increment method.}
Frequently in our mathematical analysis of expected absorption times,
we argue that some ``potential'' function of the population state increases overtime by at least a non-negligible constant rate in expectation.
We choose a potential function such that it is bounded and achieves its extreme values only when the population is homogeneous.
Thus we can compute an upper bound for the expected absorption time since the potential function will reach its extreme eventually.
Similarly, we can track the variance of a potential function overtime.
Observing the variance in the potential function allows us to compute a lower bound on the expected absorption time.

\noindent\textbf{Two types to $N$ types.}
Previous work on absorption times has concentrated on dynamics when only two types are present in the population.
Our method of analyzing the case of $N$ types involves examining the process with two types and then concluding that 
starting with $N$ types does not significantly slow down the process.
The logic is as follows:
Assign each type to one of two ``meta''-types such that the original types are roughly split between the two meta-types.
Then we run the process until one of the meta-types has fixated.
At the end of this phase there must be roughly half of the number of original types remaining.
We recursively repeat this meta-type assignment with the remaining individuals.
Since the number of types is reduced by a half after each phase,
the number of phases is logarithmic in the population size.

\noindent\textbf{Diversity.}
We measure diversity by the number of distinct types in the population.
A population with $N$ types has maximum diversity and
a population with one type has no diversity.

\noindent\textbf{Characteristic curves.}
At each time step of the evolutionary process, we can compute the diversity of the population.
Diversity can only decrease overtime since no new types arise in the population.
The characteristic curve of a population structure maps time to the expected diversity at that time
in the population.

\noindent\textbf{Properties of small graphs.}
Using the \texttt{nauty} software suite, we calculated the exact expected absorption times for all undirected graphs of
sizes $N \leq 8$ under birth-death and death-birth updating \cite{mckey1990nauty}.
We created a linear system similar to Equation~\eqref{eq:abs-time-system} but we removed symmetries.
Given that location $i$ is occupied by one of $N$ types at any given time,
there are $N^N$ possible states $\mathbf x$ in Equation~\eqref{eq:abs-time-system};
for $N=8$ there are $8^8 \approx 1.6\times 10^7$ possible states.
However, each type has the same relative fitness and does not mutate.
Thus an unlabeled partition of the population based on the locations of the types suffices to create a system for the absorption time.
Each state of our reduced system corresponds to a partition of $\{1,\ldots,N\}$.
Thus the number of equations in our system is the $N$th Bell number;
the $8$th Bell number is 4140.
For $\Omega$, a partition of $\{1,\ldots,N\}$, our reduced system is
\begin{equation}
    \tau_\Omega =
    \begin{cases}
        0&\text{if }|\Omega| = 1, \\
        1 + \sum_{\Omega'} p_{\Omega\rightarrow\Omega'}\cdot \tau_{\Omega'}&\text{otherwise}
    \end{cases}
\end{equation}

\noindent\textbf{Properties of large graphs.}
We conducted simulations of both birth-death and death-birth updating to estimate expected absorption times.
We examined graphs of sizes up to $N=100$.
For undirected graphs we estimated the expected absorption times for complete, cycle, double star, and star graphs.
We conducted numerical calculations and simulations on a high-performance remote computing cluster and across multiple nodes in order to speed up our data collection.

\noindent\textbf{Data availability.}
All simulations and numerical calculations were performed using Python 3.11.
Our code is available at \url{https://github.com/harvard-evolutionary-dynamics/diversity-time/}
and \texttt{\href{https://zenodo.org/records/14673094}{10.5281/zenodo.14673093}}.

~\\[0.4cm]
\noindent
{\bf Acknowledgments.} \\
This project was completed in part with the Illinois Combinatorics Lab for Undergraduate Experience (ICLUE).
This manuscript was posted on a preprint server: \url{https://arxiv.org/abs/2503.09841}.

\noindent
{\bf Author contributions.}\\
All authors conceived the study, performed the analysis, discussed the results, and wrote the manuscript.\\

\noindent
{\bf Competing interests.}\\
The authors declare no competing interests.\\



{\setlength{\bibsep}{0\baselineskip}
\bibliographystyle{unsrt}
\bibliography{sources}

}

}
\end{document}


\maketitle

\tableofcontents

\section{Model}\label{section:model}
We consider the multi-type Moran process \cite{goldberg2024parameterised}.
Let $\Graph=(\Vertices,\Edges)$ be a simple connected graph with vertex set $\Vertices \equiv \Set{1,\ldots,N}$ and edge set $\Edges$.
Unless otherwise stated, graphs are undirected, connected, and have
neither self-loops nor multiple edges between the same nodes.
Let $\Gamma\colon \Vertices\to 2^{\Vertices}$ be such that $\Gamma(u) = \{v \mid (u, v)\in \Edges\}$ for each $u\in \Vertices$.
For a set of types $\tau$, let $f\colon \tau\to\mathbb{Q}_{\geq 1}$ be a fitness function.
Let $\Omega$ be the set of all functions that maps elements of $V$ to elements of $\tau$.
In other words, $\Omega$ is the set of all possible \emph{configurations of types} on the vertices of $\Graph$.
We sometimes refer to an element of $\Omega$ as a \emph{configuration} for brevity.
With respect to $G$, $\tau$, $f$, and $X_0\in\Omega$, the \textit{multi-type Moran process} is a discrete-time stochastic process $(X_t)_{t\geq0}$
such that for each $t=0,1,2,\ldots$,
\begin{enumerate}
    \item $X_{t+1} \in \Omega$, and
    \item $X_{t+1}$ is such that $X_{t+1}(D_{t+1}) = X_t(B_{t+1})$ and $X_{t+1}(u)=X_t(u)$ for all $u\in \Vertices\setminus\{D_{t+1}\}$; denote this operation $R(X_t, D_{t+1},B_{t+1})=X_{t+1}$
\end{enumerate}
where for a Birth-death process, we have
\begin{enumerate}
    \item $B_{t+1} = u$ with probability $f(X_t(u))/\sum_{x\in V} f(X_t(x))$ for each $u\in V$, and
    \item $D_{t+1} = v$ with probability $1/\deg(B_{t+1})$ for each $v\in \Gamma(B_{t+1})$.
\end{enumerate}
and for the death-Birth process, we have
\begin{enumerate}
    \item $D_{t+1} = v$ with probability $1/|V|$ for each $v\in V$, and
    \item $B_{t+1} = u$ with probability $f(X_t(u))/\sum_{x\in \Gamma(D_{t+1})} f(X_t(x))$ for each $u\in \Gamma(D_{t+1})$.
\end{enumerate}
In this work, we will consider constant fitness functions unless otherwise stated.
When the fitness function is constant, we will denote Birth-death (Bd) and death-Birth (dB) processes as birth-death (bd) and death-birth (db) processes, respectively.
We will also only consider the case where both $|\tau| = |V|$ and $X_0(V) = \tau$ unless otherwise stated.
For a graph $\Graph=(\Vertices,\Edges)$ and $U\subseteq\Vertices$, let $\phi_\Graph(U):= \sum_{u\in U}\frac{1}{\deg(u)}$ where $\deg\colon\Vertices\to\mathbb Z_{\geq 1}$ is the degree of $u$
in $\Graph$.
\cite{lieberman2005evolutionary}.

Consider a connected undirected graph where initially
every vertex is occupied by a distinct type.
The \emph{diversity time} for a particular update rule
is the expected number of steps the process takes until only
one type remains.
We call a step of the process \emph{active}
if the placements of the wild-types or mutants have changed
as a result of this step.

\input{./si/proofs/well_mixed}

\input{./si/proofs/cycle.tex}

\input{./si/proofs/star.tex} 

\input{./si/proofs/long_lemma}

\input{./si/proofs/double_star.tex} 

\input{./si/proofs/barbell.tex} 

\input{./si/proofs/upper_lower_bounds.tex} 
 
\input{./si/proofs/directed_graph.tex} 

\bibliography{si}
\bibliographystyle{unsrt}

\clearpage
\newpage
\section{Additional Figures}

\renewcommand{\figurename}{\textbf{Figure}}
\renewcommand{\thefigure}{\textbf{S\arabic{figure}}}
\setcounter{figure}{0}

\begin{figure}[h]
    \centering
    \includegraphics[width=\linewidth]{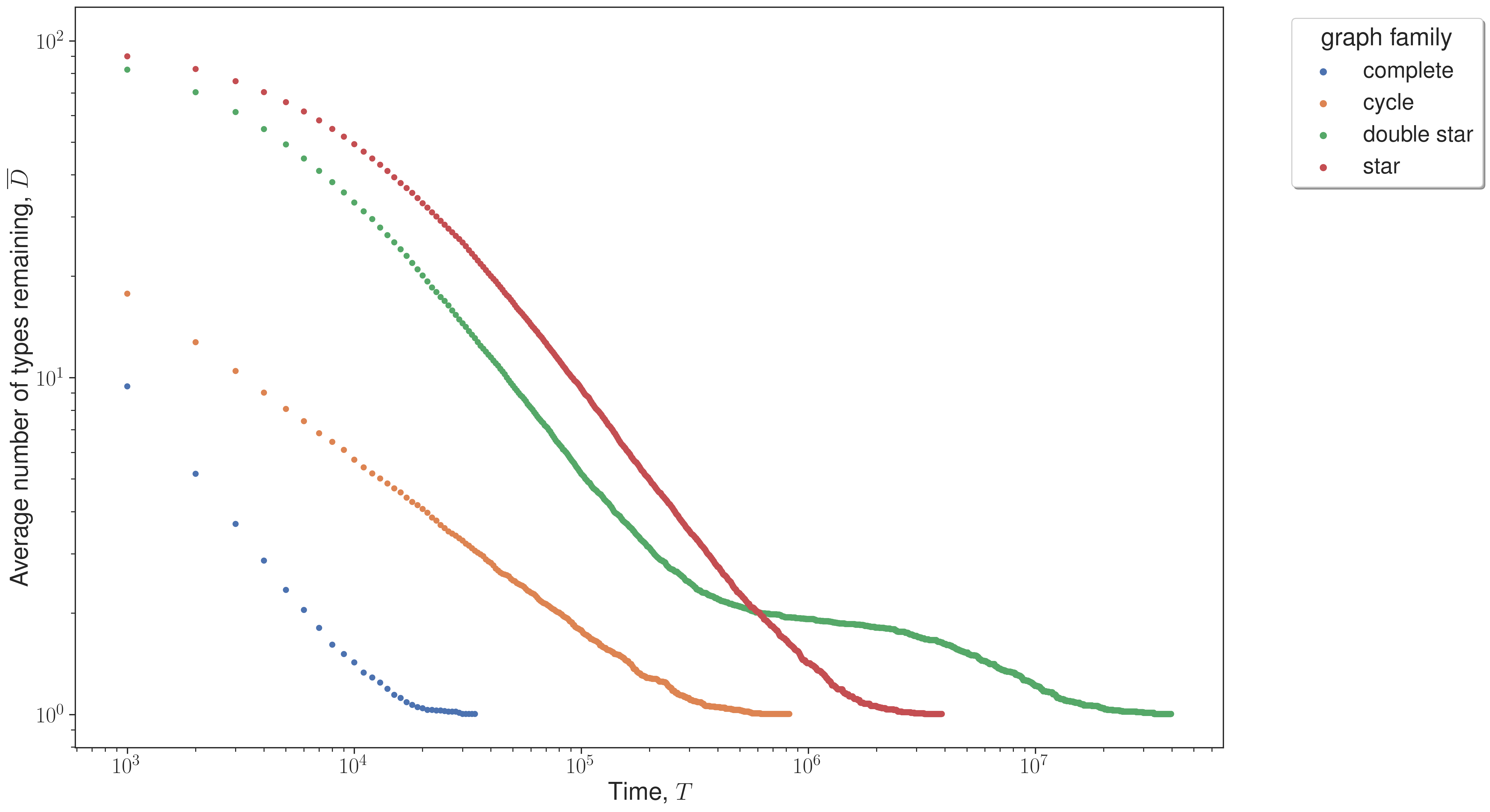}
    \caption{%
        Results of average number of types remaining in the population, $\overline{D}$, at time $T$, averaged over $250$ simulations
        of birth-death updating
        per graph.
        Each graph has $N=100$ vertices.
        The plot is on a log-log scale.
        If $\overline{D}=1$ at a particular time $T$, no dot is drawn.
        Only times that are multiples of $10^3$ (excluding $0$) are plotted.
    }
    \label{fig:characteristic-curves-N100}
\end{figure}

%% file: si/proofs/well_mixed.tex
\section{Well-mixed}\label{sec:complete}
Consider running the multi-type process on a complete graph with self-loops and $n$ vertices;
 this is known as a well-mixed population.
Due to the symmetries of a complete graph, we can track each state of this multi-type process by keeping track of solely the abundance of each type.
We use common notation in combinatorics to track such abundances; see \cite{stanley2011enumerative}.

Let $\lambda\equiv(\lambda_1,\ldots,\lambda_\ell)$ such that
$\lambda_i \in \mathbb Z_{>0}$ for each $i\in [\ell]$,
$\lambda_i \geq \lambda_{i+1}$ for each $i\in[\ell-1]$,
and $\sum_{i=1}^\ell \lambda_i = n$.
Thus $\lambda$ is an \emph{integer partition of $n$} of \emph{length} $\ell$.
We use the notation $\lambda\vdash n$ to denote that $\lambda$ is an integer partition of n.
If $\lambda\vdash n$ is of length $\ell$ and $i\not\in[\ell]$, we let $\lambda_i=0$.

Notice that this multi-type Moran process is equivalent to the following partition walk.
\begin{definition}[Frequency-dependent partition walk] 
    Given $\lambda\vdash n$ of length $\ell$, a step in the walk from $\lambda$ to $\lambda''\vdash n$ consists of following operations:
    \begin{enumerate}
        \item (birth) Choose $r\in [\ell]$ with probability $\lambda_r/n$;
        \item (death) Choose $r'\in [\ell]$ with probability $\lambda_{r'}/n$;
        \item (replacement)
        \begin{enumerate}
            \item Construct a partition $\lambda'$ by setting $\lambda'_r = \lambda_r-1$, $\lambda'_{r'}=\lambda_r+1$, and setting $\lambda'_i = \lambda_i$ for the remaining indices;
            \item Construct a partition $\lambda''$ by sorting the values of $\lambda'$ in decreasing order and then removing any indices of $\lambda'$ that have value $0$;
        \end{enumerate}
        Denote this operation $R(\lambda,r,r')=\lambda''$.
    \end{enumerate}
\end{definition}
Notice that the singleton partition is the sole absorbing state of this walk.
Denote hitting time of the frequency-dependent partition walk
from any partition $\lambda\vdash n$ to the singleton partition $(n)$ as $t^{\text{hit}}_\lambda$.
\begin{theorem}\label{thm:well-mixed-formula}
    Let $\lambda \vdash n$ have length $\ell$.
    Then 
    \begin{equation}\label{eq:well-mixed-formula}
        \E{t^{\text{hit}}_\lambda} = n^2 - n - \sum_{i=1}^\ell\sum_{k=0}^{\lambda_i-1}\frac{(n+\lambda_i-2k)k}{n-k}.
    \end{equation}
\end{theorem}

\begin{figure}[ht!]
    \centering
\includegraphics[width=0.25\linewidth]{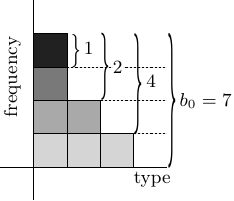}
    \caption{One intuitive way to think about the formula for the complete graph. Here $n=7$ and $\lambda=(4,2,1)$, thus $(b_0,b_1,b_2,b_3)=(7,4,2,1)$. Then $\E{t^{\text{hit}}_\lambda} =7\cdot (7-\frac77-\frac46 -\frac25-\frac14)=32.78\overline{3}$.}
    \label{fig:intuition}
\end{figure}

One intuitive way to think about the formula is this (see \cref{fig:intuition}):
Think about the ``histogram'' of the partition $\lambda$ and for each $h\ge 0$,
denote by $b_h$ the number of boxes above the line $y=h$. (In particular, $b_0=n$.)
An explicit formula for $b_h$ is $\sum_{i} \max(\lambda_i-h,0)$.
Then \[
\E{t^{\text{hit}}_\lambda} = n\cdot\left(n - \sum_{h=0}^{n-1} \frac{b_h}{n-h}\right).\]

We first prove the intuition behind the formula in \cref{fig:intuition} and then prove \cref{thm:well-mixed-formula}.
\begin{lemma}
    Let $\lambda\vdash n$ have length $\ell$.
    Then
    \begin{equation}
        n^2 - n - \sum_{i=1}^\ell\sum_{k=0}^{\lambda_i-1}\frac{(n+\lambda_i-2k)k}{n-k}
        =n\cdot\left(n - \sum_{h=0}^{n-1} \frac{b_h}{n-h}\right).
    \end{equation}
\end{lemma}
\begin{proof}
    Using the notation $\mathbbm{1}[P]$ to denote the indicator function of the predicate $P$,
    we interchange summation on the left hand side to get
    \begin{align}
       n^2 - n - \sum_{i=1}^\ell\sum_{k=0}^{\lambda_i-1}\frac{(n+\lambda_i-2k)k}{n-k} 
       &= n^2 - n - \sum_{k=0}^{n-1}\sum_{i=1}^\ell\frac{(n+\lambda_i-2k)k\cdot\mathbbm{1}[k < \lambda_i]}{n-k} 
    \end{align}
    Notice that $n+\lambda_i-2k = (n-k) + (\lambda_i-k)$.
    This means that
    \begin{align}
       n^2 - n - \sum_{k=0}^{n-1}\sum_{i=1}^\ell\frac{(n+\lambda_i-2k)k\cdot\mathbbm{1}[k < \lambda_i]}{n-k} 
       &= n^2 - n - \sum_{k=0}^{n-1}\sum_{i=1}^\ell\frac{k(n-k)}{n-k}\cdot\mathbbm{1}[k < \lambda_i] 
       -\sum_{k=0}^{n-1}\sum_{i=1}^\ell\frac{k(\lambda_i-k)}{n-k}\cdot\mathbbm{1}[k < \lambda_i]   \\
       &= n^2 - n - \sum_{k=0}^{n-1}\sum_{i=1}^\ell k\cdot\mathbbm{1}[k < \lambda_i] 
       -\sum_{k=0}^{n-1}\sum_{i=1}^\ell\frac{k\cdot\max(\lambda_i-k, 0)}{n-k}   \\
       &= n^2 - n - \sum_{k=0}^{n-1}k\sum_{i=1}^\ell \mathbbm{1}[k < \lambda_i] 
       -\sum_{k=0}^{n-1}\frac{k}{n-k}\sum_{i=1}^\ell\max(\lambda_i-k, 0)   \\
       &= n^2 - n - \sum_{i=1}^\ell\sum_{k=0}^{\lambda_i-1} k
       -\sum_{k=0}^{n-1}\frac{k\cdot b_k}{n-k}. \label{eq:reduced-rhs}
    \end{align}
    Now notice that
    \begin{align}
        \sum_{k=0}^{n-1}\frac{k\cdot b_k}{n-k}
        &=  
        \sum_{k=0}^{n-1}\frac{n\cdot b_k}{n-k}
        - \sum_{k=0}^{n-1}\frac{(n-k)\cdot b_k}{n-k} \\
        &=  
        n\cdot\sum_{k=0}^{n-1}\frac{b_k}{n-k}
        - \sum_{k=0}^{n-1} b_k.
    \end{align}
    We also have that
    \begin{align}
       \sum_{k=0}^{n-1} b_k 
       = \sum_{k=0}^{n-1}\sum_{i=1}^\ell \max(\lambda_i-k, 0)
       = \sum_{k=0}^{n-1}\sum_{i=1}^\ell \sum_{h=k}^{\lambda_i-1} 1
       = \sum_{i=1}^\ell \sum_{k=0}^{\lambda_i-1} \lambda_i-k.
    \end{align}
    Thus \Cref{eq:reduced-rhs} becomes
    \begin{align}
        n^2 - n - \sum_{i=1}^\ell\sum_{k=0}^{\lambda_i-1} k
       -\sum_{k=0}^{n-1}\frac{k\cdot b_k}{n-k}
       &=  
        n^2 - n
        - \left(\sum_{i=1}^\ell\sum_{k=0}^{\lambda_i-1} k\right)
       -\left( n\cdot\sum_{k=0}^{n-1}\frac{b_k}{n-k}
        -  \sum_{i=1}^\ell \sum_{k=0}^{\lambda_i-1} \lambda_i-k\right) \\
       &= 
        n^2 - n
        - 2\cdot\left(\sum_{i=1}^\ell\sum_{k=0}^{\lambda_i-1} k\right)
        + \left(\sum_{i=1}^\ell \sum_{k=0}^{\lambda_i-1} \lambda_i\right)
       -\left( n\cdot\sum_{k=0}^{n-1}\frac{b_k}{n-k} \right) \\
       &= 
        n^2 - n
        - \left(\sum_{i=1}^\ell \lambda_i\cdot(\lambda_i-1)\right)
        + \left(\sum_{i=1}^\ell \lambda_i^2\right)
        - \left( n\cdot\sum_{k=0}^{n-1}\frac{b_k}{n-k} \right) \\
       &= 
        n^2 - n
        + \left(\sum_{i=1}^\ell \lambda_i\right)
        - \left( n\cdot\sum_{k=0}^{n-1}\frac{b_k}{n-k} \right) \\
       &= 
        n^2 - n
        + \left(n\right)
        - \left( n\cdot\sum_{k=0}^{n-1}\frac{b_k}{n-k} \right) \\
       &= 
        n^2
        - \left( n\cdot\sum_{k=0}^{n-1}\frac{b_k}{n-k} \right) \\
       &= 
        n \cdot \left( n-\sum_{k=0}^{n-1}\frac{b_k}{n-k} \right).
    \end{align}
\end{proof}
Now we prove \cref{thm:well-mixed-formula}.
\begin{proof}[Proof of Theorem \ref{thm:well-mixed-formula}]
    We encode the problem as a Markov chain.
    The states are the partitions of $n$;
    for any partitions $\lambda,\mu\vdash n$,
    the transition probability from $\lambda$ to $\mu$
    is $\sum_r\sum_{r'}\lambda_r\lambda_{r'}/n^2$
    where the sum is over all $r,r'$ such that $R(\lambda,r,r')=\mu$.
    
    For an absorbing Markov chain with states $\{s_0,\ldots,s_m\}$
    where $s_0$ is the only absorbing state, and transition probabilities
    $(q_{ij})$, the expected times to absorption $h_0,\ldots,h_m$ are the
    unique solution to the following system of linear equations:
    \begin{equation*}
        h_i =
        \begin{cases}
            0&\text{if }i=0,\\
            1+\sum_{j=0}^m q_{ij} h_j&\text{otherwise}.
        \end{cases}
    \end{equation*}
    Thus $(\E{t^{\text{hit}}_\lambda})_{\lambda\vdash n}$ is the unique
    solution $h((\lambda))_{\lambda\vdash n}$ to the system of linear equations:
    \begin{equation*}
        h(\lambda) =
        \begin{cases}
            0&\text{if }\lambda=(n),\\
            1+\sum_r\sum_{r'}\frac{\lambda_r\lambda_{r'}}{n^2}\cdot h(R(\lambda,r,r'))&\text{otherwise}.
        \end{cases}
    \end{equation*}

    First, to show \cref{eq:well-mixed-formula} holds, we consider
    when $\lambda=(n)$.
    Then we have $h(\lambda)=0$ and
    \begin{equation*}
        n^2-n-\sum_{k=0}^{n-1}\frac{(n+n-2k)k}{n-k}
        = n^2-n-\sum_{k=0}^{n-1}2k = n^2-n - (n^2-n)=0.
    \end{equation*}
    Next, consider the case when $\lambda\neq (n)$.
    Denote $T(m) := \sum_{k=1}^{m-1} \frac{(n+m-2k)k}{n-k}$ and $g(\lambda)=n^2-n-\sum_{i=1}^{\ell} T(\lambda_i)$. 
    We want to show $h(\lambda)=g(\lambda)$ for
    each $\lambda\neq (n)$.
    Further, let $c^\lambda_j:=\#\{i\in [\ell] : \lambda_i=j\}$.
    Thus $\sum_{i=1}^{\ell}T(\lambda_i) = \sum_{j=1}^n c^\lambda_j\cdot T(j)$.
    Now for any $r\neq r'$, by plugging in $g(\lambda)$ into $h(\lambda)$
    for each $\lambda\vdash n$, we can write
    \begin{align*}
        g(R(\lambda,r,r'))
        &= \left(n^2-n-\sum_{j=1}^nc_j\cdot T(j)\right)-\left(T(\lambda_r)+T(\lambda_{r'})\right) + \left(T(\lambda_r-1)+T(\lambda_{r'}+1)\right) \\
        &= g(\lambda)-\left(T(\lambda_r)+T(\lambda_{r'})\right) + \left(T(\lambda_r-1)+T(\lambda_{r'}+1)\right).
    \end{align*}
    When $r=r'$, we have $g(R(\lambda,r,r'))=g(\lambda)$.
    Hence, we need to show
    \begin{equation}\label{eq:need-to-show}
    \sum_r\sum_{r'\neq r} \frac{\lambda_r\lambda_{r'}}{n^2}\left(-T(\lambda_r)-T(\lambda_{r'})+T(\lambda_r-1)+T(\lambda_{r'}+1)\right) = 1.
    \end{equation}
    
    First, we have
    \begin{align*}
        T(m+1)-T(m)
        &= \sum_{k=1}^m\frac{(n+m-2k+1)k}{n-k} - \sum_{k=1}^{m-1}\frac{(n+m-2k)k}{n-k} \\
        &= \frac{(n-m+1)m}{n-m} + \sum_{k=1}^{m-1}\frac{(n+m-2k+1)k}{n-k}-\frac{(n+m-2k)k}{n-k} \\
        &= \frac{(n-m+1)m}{n-m} + \sum_{k=1}^{m-1}\frac{k}{n-k} \\
        &= m+\frac{m}{n-m} + \sum_{k=1}^{m-1}\frac{k}{n-k} \\
        &= m+ \sum_{k=1}^{m}\frac{k}{n-k}.
    \end{align*}
    Let $U(m) := \sum_{k=1}^{m-1}\frac{k}{n-k}$.
    Thus the left hand side of \cref{eq:need-to-show} is
    \begin{align*}
        &\sum_r\sum_{r'\neq r} \frac{\lambda_r\lambda_{r'}}{n^2}
        \left(U(\lambda_{r'}+1)-U(\lambda_r)+\lambda_{r'}-\lambda_r+1\right) \\
        &= \sum_r\sum_{r'>r}\frac{\lambda_r\lambda_{r'}}{n^2}
         \left(
         U(\lambda_{r'}+1)-U(\lambda_r)+\lambda_{r'}-\lambda_r+1
         +U(\lambda_{r}+1)-U(\lambda_{r'})+\lambda_{r}-\lambda_{r'}+1
         \right) \\
        &= \sum_r\sum_{r'>r}\frac{\lambda_r\lambda_{r'}}{n^2}
         \left(
         1 + \frac{\lambda_r}{n-\lambda_r}
         +1 + \frac{\lambda_{r'}}{n-\lambda_{r'}}
         \right) \\
        &= \sum_r\sum_{r'>r}\frac{\lambda_r\lambda_{r'}}{n^2}
         \left(
         \frac{n}{n-\lambda_r}+\frac{n}{n-\lambda_{r'}}
         \right) \\
        &= \sum_r\sum_{r'>r}\frac{\lambda_r\lambda_{r'}}{n}
         \left(
         \frac{1}{n-\lambda_r}+\frac{1}{n-\lambda_{r'}}
         \right) \\
        &= \frac1n\sum_r\frac{\lambda_r}{n-\lambda_r}\sum_{r'>r}\lambda_{r'}
         + \frac1n\sum_r\frac{\lambda_r}{n-\lambda_r}\sum_{r'<r}\lambda_{r'} \\
        &= \frac1n\sum_r\frac{\lambda_r}{n-\lambda_r}\sum_{r'>r}\lambda_{r'}
         + \frac1n\sum_r\frac{\lambda_{r'}}{n-\lambda_{r'}}\sum_{r<r'}\lambda_{r} &\text{(by relabeling)} \\
        &= \frac1n\sum_{r}\frac{\lambda_r}{n-\lambda_r}\sum_{r'\neq r} \lambda_{r'}.
    \end{align*}
    Finally, we use the fact that $\lambda_1 + \cdots + \lambda_\ell=n$
    to obtain
    \begin{equation*}
        \frac1n\sum_{r}\frac{\lambda_r}{n-\lambda_r}\sum_{r'\neq r} \lambda_{r'}
        = \frac{1}{n}\sum_{\lambda_r}{n-\lambda_r}(n-\lambda_r)
        =\frac{1}n\sum_r\lambda_r
        =\frac1n\cdot n
        =1.
    \end{equation*}
    Therefore $h(\lambda)=g(\lambda)$ for each $\lambda\vdash n$, and thus $\E{t^{\text{hit}}_\lambda}=g(\lambda)$ for each $\lambda\vdash n$. 
\end{proof}

\begin{corollary}\label{cor:well-mixed-fully-diverse}
    Let $(1,\ldots,1)\vdash n$.
    Then $\E{t^{\text{hit}}_{(1,\ldots,1)}} = n^2-n$.
\end{corollary}
\begin{proof}
    By \cref{thm:well-mixed-formula},
    \begin{equation*}
        \E{t^{\text{hit}}_{(1,\ldots,1)}}
        =n^2-n-\sum_{i=1}^n\sum_{k=0}^{0} \frac{(n+1-2k)k}{n-k}=n^2-n.
    \end{equation*}
\end{proof}

\begin{corollary}
    The expected absorption time of the multi-type
    process on a complete graph with no self-loops and $n$ vertices is $(n-1)^2$ when starting with $n$
    different types.
\end{corollary}
\begin{proof}
    Let $G=(V,E+\{(u,u)\mid u\in V\})$ be a complete graph with self-loops on $n$ vertices.
    At each step, regardless of the configuration,
    there is a $1-1/n$ probability a birth occurs along
    $E$.
    Otherwise, the configuration does not change.
    Thus by linearity of expectation, the expected absorption time for $G$ is the same as the expected absorption time for $G'=(V,E)$ where
    each step has weight $(1-1/n)^{-1}$ (expectation of a geometric random variable with success probability $1-1/n$).
    So from \Cref{cor:well-mixed-fully-diverse} we get
    that the expected absorption time for $G'$ is
    \begin{equation}
        (1-1/n) \cdot (n^2-n)
        = ((n-1)/n) \cdot n(n-1)
        = (n-1)^2.
    \end{equation}
\end{proof}

%% file: si/proofs/cycle.tex
\section{Cycles}

We consider the multi-type process on a cycle and show the exact expected time from any configuration.
The result uses a potential function that tracks the expected absorption time and implies $\Theta(n^3)$ for the diversity time.
We present the proof for completeness, the proof can also be derived as a special case of~\cite{broom2010evolutionary}.

\begin{observation}\label{obs:conti}
    Since every type starts with only one individual and reproductions happen only on the boundaries of contiguous segments of a type,
    each type always occupies a contiguous segment of the cycle.
\end{observation}

Similar to our analysis of well-mixed populations,
we use $\lambda$, an \emph{integer partition of $n$} of \emph{length} $\ell$.
This time, $\lambda$ does not encode the whole configuration on the graph, there are nonisomorphic configurations that have the same $\lambda$.
However, $\lambda$ and \Cref{obs:conti} are enough to track the expected number of steps.

\begin{theorem}
    Given $\lambda$ on a cycle where every type forms a contiguous segment of the cycle, we have
    \[
        \E{t^{\text{hit}}_\lambda} = \frac{(n-1)n(n+1)}6 -\sum_{h=0}^{n-1} b_h\cdot h,
    \]
    where $b_h := \sum_{i \in [\ell]} \max(0, \lambda_i - h)$.
\end{theorem}
\begin{proof}
    Given a configuration $\lambda$, we use the potential function
\begin{equation*}
    \phi(\lambda) =  n^2 - n - \sum_{i=1}^\ell\sum_{k=0}^{\lambda_i-1}\frac{(n+\lambda_i-2k)k}{n-k}
\end{equation*}
which is the expected absorption time on a complete graph with $n$ vertices.

We show that given the potential and $\lambda$, the expected change is $-1$ unless the whole cycle is occupied by only one type.
Since the potential is $0$ if the cycle is occupied by one type, this implies that the potential tracks the expected number of steps.

We can substitute the bd or db process by selecting a random edge and then selecting the direction of reproduction.
This change keeps the same replacement probabilities.

We examine the potential change conditioned on the fact that an edge between two different individuals $i$ and $j$ is selected.
Suppose that $\lambda_i \ge \lambda_j$ and $i$ replaced individual $j$.
Then $b_h$ changes to $b_h'$ and $b_h' - b_h = 1$ for all $h$ from $\lambda_j$ to $\lambda_i$.
Similarly, when individual $j$ replaces $i$, the difference in $b_h'-b_h$ is $-1$ for all $h$ from $\lambda_j+1$ to $\lambda_i-1$.

Since the edge between $i$ and $j$ has the same probability of being selected in both directions, we have the expected change in the potential
\[
    -\frac{1}{2}\lambda_i-\frac{1}{2}\lambda_j\,.
\]

We have this change in potential for any pair of neighboring types.
There are $|\ell|$ neighboring types, conditioned that an edge between them is selected, we have the change in potential
\[
    \frac{1}{|\ell|}\sum_{i \in [\ell]} -\lambda_i = -\frac{n}{|\ell|}\,.
\]
Finally, an edge between two individuals is selected with probability $\frac{|\ell|}{n}$, which gives the change in potential $-1$ in one step.
\end{proof}

\section{Paths}
The path is created from the cycle by removing a link between vertex $v_1$ and $v_n$.
We use the potential for the cycle with minor modifications and prove bound $\Theta(n^3)$ for the diversity time.

\begin{theorem}\label{thm:path}
    The diversity time on a path is $\Theta(n^3)$.
\end{theorem}
\begin{proof}
    First, we show the lower bound on the diversity time.
    We show that the expected fixation time is $\Omega(n^3)$ for two types, each occupying a continuous half of the path.
    This configuration can be simulated by the unbiased one-dimensional Markov Chain where the absorbing states are at distances $\Theta(n)$ from the starting state.
    The expected time on this Markov Chain is $\Theta(n^2)$ from Proposition 2.1 of~\cite{levin2017markov};
    however, we have an $n$ times slowdown because most of the time individuals not on the boundary reproduce.

    Observe that this also shows that for two types, the expected time is $\calO(n^3)$.
    In other words, the time until one type disappears is bounded by $\calO(n^3)$.
    This means that the diversity time for a constant number of types is also $\calO(n^3)$.

    For many types, we use the same potential as in the cycle but we change the process such that in every configuration, the potential decreases by at least $-1$.

    Any type that at any time occupies $v_1$, $v_2$ (neighbor of $v_1$), $v_{n-1}$ (neighbor of $v_n$), or $v_n$ is treated as being type $0$.
    Type $0$ is an amalgamation of at most $4$ types, two from each end of the path.
    In the beginning, type $0$ consists of four types, and any type that becomes type $0$ needs to claim $v_2$ or $v_{n-1}$, but if this happens, there can be only one other type in $v_1$ or $v_n$.

    When no type is changed to type $0$, the potential decreases as in the cycle.
    Our changes ensure that the type $0$ occupies both vertices of degree $1$ and their neighbors.
    So, the only vertices that have different replacement probabilities than the cycle are occupied by type $0$, and the replacement of the same type does not change the potential.

    When another type becomes type $0$, then the potential decreases more than in the cycle, it is equivalent to merging two types.

    This means the potential decreases by at least $1$ every step.
    Therefore the expected time until only one type remains is $\calO(n^3)$.
    The remaining type is type $0$, which was created by fusing of at most $4$ different types.
    However, we know that in that case, the expected time is at most $\calO(n^3)$.

    The expected time on the path is at least $\Omega(n^3)$ even for two types, the time until there are only constantly many types requires $\calO(n^3)$ steps and finally getting to one type requires $\calO(n^3)$ steps.
    Therefore the expected time on path is $\Theta(n^3)$.

\end{proof}

%

%% file: si/proofs/star.tex
\section{Stars}

\subsection{bd updating}
We use the results from \Cref{sec:complete} to show a cubic bound for stars.
We consider a star with $n$ leaves and thus $N = n+1$ total vertices.

\begin{theorem}\label{thm:bd-star-upper}
    The diversity time for a star in bd updating is $\calO(N^3)$.
\end{theorem}
\begin{proof}
Note that every configuration of the bd
process on a graph $G=(V,E)$ with $|V|=n$ can be represented by
a partition of $\{1,\ldots,n\}$.
Let $\Pi$ be all partitions of the $n$
leaves of a star.
We define a potential function $\phi\colon\Pi\to \Real$
that maps a given partition $\lambda$ of $\{1,\ldots,n\}$
to the expected absorption time on a complete graph with $n$ vertices
with initial configuration $\lambda$.

We will track only the reproductions of individuals on the central vertex and the configuration in the leaves of the star.
Any time the center individual reproduces (which happens with probability $1/N$), two things could happen in the previous round:
\begin{enumerate}
    \item[(A)] an individual from a leaf replaced the center individual (with probability $1-1/N$), or
    \item[(B)] the center individual reproduced.
\end{enumerate}
Whenever Situation (A) happens, the probability distribution of the center corresponds to $\lambda$ up to reordering.
Therefore the transition probabilities are the same as in the complete graph:
in the previous round, a randomly chosen individual replaced the center.
Whenever Situation (B) happens, we examine the probabilities and in some cases, we select the most pessimistic replacement that increases the potential $\phi$ the most.
It is the event where the least prevalent type in $\lambda$ replaces one individual of the most prevalent type in $\lambda$.
Even with these changes, we show that, in every configuration, the potential decreases by at least a constant in expectation.
This will give the bound $\calO(N^2)$ for the number of reproductions of the center from Theorem 7 of \cite{kotzing2019first}.
Since the center reproduces with probability $1/N$, counting all reproductions including leaves replacing the center gives the total number of steps $\calO(N^3)$.

Now, for the configuration $\lambda\in\Pi$ on the leaves of the star, we have the following potential.
\begin{equation*}
        \phi(\lambda) = n^2 - n - \sum_{i=1}^\ell\sum_{k=0}^{\lambda_i-1}\frac{(n+\lambda_i-2k)k}{n-k}.
\end{equation*}

Let $\lambda^{(i)}$ be the configuration of the leaves after $i$ steps of the process.
Suppose the center reproduces for the $i$th step.
Then
\begin{itemize}
    \item with probability $1-1/N$, an individual from a leaf reproduced
    into the center for the $(i-1)$th step.
    Then the potential changes by \[
        \phi(\lambda^{(i)})-\phi(\lambda^{(i-1)})
        =-1
    \]
    since this is one step in the process for the
    complete graph with self-loops.
    The expected absorption time (which is the potential function) will decrease by 1.

    \item
    with probability $1/N$, the center individual gave birth for the $(i-1)$th step.
    \begin{itemize}
        \item with probability $1-1/N$, a leaf individual gave birth for the $(i-2)$th step.
        We know that $\lambda^{(i-1)}$ and $\lambda^{(i)}$ can only differ in at most two locations.
        The probability of selecting any of the differing indices is at most $2/n$ since there are $n$ leaves.
        Observe that the biggest possible change in the potential $\phi$ is \[ n-1+ \sum_{k=1}^{n-1}\frac{k}{n-k} \le n H_n \le n \log n. \]
        This happens when a type that occupies all but one leaf is replaced.
        Specifically, for a given configuration $\lambda$, the biggest increase in potential happens when the least prevalent type replaces the most prevalent type.
        Thus with probability at most $2/n$ the potential function has increased by at most $n\log n$ and with remaining probability has decreased by $1$.
        
        \item with probability $1/N$, the center individual gave birth for the $(i-2)$th step.
        By similar logic, the potential can only increase by $n\log n$ at each step.
    \end{itemize}
\end{itemize}

Putting the pieces together, we get
\begin{align*}
    E[\phi(\lambda^{(i)})-\phi(\lambda^{(i-1)})\mid \lambda^{(i-1)}]
    &\leq
    (1-1/N)\cdot (-1) \\
    &+ (1/N) \cdot (1-1/N)\left[(2/n) \cdot n\log n + (1-2/n)\cdot (-1) \right] \\
    &+ (1/N)\cdot (1/N) \cdot n\log n \\
    &\leq
    -1+1/n \\
    &+ (1/n) \cdot \left(2\log n -1+2/n \right) \\
    &+ (\log n)/n \\
    &= -1 + (3\log n)/n + 2/n^2 \\
    &= -1 + o(1).
\end{align*}
The completion of the proof follows from the Theorem 7 of \cite{kotzing2019first}.


%
%
%
\end{proof}

\subsection{db updating}

\begin{theorem}
    The diversity time on a star in db updating is bounded by $\calO(N\log N)$.
\end{theorem}
\begin{proof}
    Let $n$ be the number of leaves of the star
    and $N=n+1$ be the total number of vertices.
    We show that with high probability, the individual in the center spreads over the whole graph in $n$ active steps.
    
    Let $i$ be the number of individuals whose types differ from the type of the center individual.
    Two possibilities exist for an active step:
    \begin{enumerate}
        \item[(A)] $i$ decreases, or
        \item[(B)] the center individual is replaced.
    \end{enumerate}
    Event~(A) happens if one of the individuals different from the center dies;
    the probability of this event is $i/N$.
    Event~(B) happens when the center dies (probability $1/N$) and is replaced by a different individual (probability $i/n$).
    Thus in one active step, the probability of losing the center is
    \[
        \frac{\frac{1}{N}\frac{i}{n}}{\frac{1}{N}\frac{i}{n} + \frac{i}{N}} = 
        \frac{\frac{1}{n}}{\frac{1}{n} + 1} = \frac{1}{N}\,.
    \]
    One type must fixate after at most $n$ active steps if the center was never replaced;
    this happens with probability
    \[
        \left(1 - \frac{1}{N}\right)^{n} \ge 1/e\,.
    \]
    
    Now, we compute the maximal expected number of all steps until the center fixates or is replaced.
    Having $i$ different individuals different from the center, the probability of an active step is at least $i/N$.
    That means to have all $n$ leaves to be the same type as the center or have the center replaced,
    we have expected time at most
    \[
        \sum_{i = 1}^n \frac{N}{i} = N \cdot H_n. 
    \]
    Then the total expected time until absorption, $T_N$, satisfies.
    \[ T_N \leq (N\cdot H_n) + (1-1/e)\cdot T_N. \]
    Thus $T_N \leq eN\cdot H_n \leq \calO(N\log N)$.
\end{proof}

%% file: si/proofs/long_lemma.tex
\section{General Lemma for Lower Bounds}

In this section, we prove a general lemma about lower bounds for absorption times.
We consider a process where one of the types cannot be replaced (is invincible) when occupying certain vertices.
Then we show the relationship between this process and the original process.
The result is general and holds for bd, db, and all dynamics where the replacement vertices do not depend on the exact neighborhood configurations.

\paragraph{Notation.}
Let the type $a$ occupy a set of vertices $A$.
We denote the fixation probability of type $a$ occupying the set $A$ by $f(a, A)$.
Let $T$ be a random variable that denotes the absorption time of the process when type $a$ initially only occupies $A$.

\paragraph{The invincible process.}
Let the \emph{invincible process} for set $A$ be similar to the original process,
but where individuals residing on vertices in $A$ cannot be replaced.
Let $T_A$ be a random variable that denotes the absorption time of
this invincible process for set $A$ where type $a$ initially only occupies $A$.

\begin{lemma}\label{lemma:general_lower}
  Let $G=(V,E)$ and $A\subseteq V$.
  For bd and db updating, we have
  \[
    E[T] \ge \frac{f(a,A)^2}{4} E[T_A]\,.
  \]
\end{lemma}
\begin{proof}
    We have 
    \begin{equation}\label{eq:general-lower-master}
        E[T] = f(a,A)E[T\mid a \text{ fixates}] + (1-f(a,A))E[T\mid a \text{ becomes extinct}]\,.
    \end{equation}
    Let $E_f$ denote the expected absorption time of the invincible process for $A$ conditioned on the event the trajectory is among the shortest $f(a,A)$-proportion of the possible invincible trajectories.
    We will first show
    \[
        E[T\mid a \text{ fixates}] \ge E_f \,.
    \]

    Consider running the original process $\mathcal P_{\emptyset}$
    and the invincible process $\mathcal P_A$ with the same
    randomness and where type $a$ initially only occupies $A$.
    Suppose $\mathcal P_{\emptyset}$ absorbs with type $a$ fixated.
    Since individuals from $A$ are not replaced, $\mathcal P_A$ must have
    absorbed with type $a$ fixated but in no more steps than $\mathcal P_{\emptyset}$.
    That means in $\mathcal P_A$, type $a$ occupies a superset of vertices 
    at each step compared to the vertices that type $a$ occupies in $\mathcal P_\emptyset$.
    So we have \[
        E[T\mid a \text{ fixates}] \geq E[T_A \mid a\text{ fixates}]
        \geq E_f.
    \]

    Now, we compare $E[T_A]$ with $E_f$.
    We run $\mathcal P_A$ for $2 E_f$ steps.
    From Markov's inequality,
    \[
        P\left[T_A > 2 E_f \bigm| \mathcal P_A \text{ is among shortest } f(a,A)\text{-proportion of invincible trajectories}\right]
        \leq \frac{E_f}{2 E_f} = 1/2.
    \]
    Thus \[ P[T_A \leq 2E_f] \geq f(a, A)/{2}. \]
    So we have that at least $f(a, A)/{2}$ proportion of invincible process absorbs within $2E_f$ steps.
    If $\mathcal P_A$ does not absorb within $2E_f$ steps, type $a$ occupies vertices from $A$ and possibly some other vertices.
    From this configuration, the expected time is at most $E[T_A]$, since the set of vertices occupied by $a$ is a superset of $A$.
    That means
    \[
        E[T_A] \le 2 E_f + \left(1-\frac{f(a,A)}{2}\right)E[T_A]\,.
    \]
    That means $E[T_A] \le \frac{4}{f(a,A)}E_f$.
    Combining the bound on $E[T_A]$ with Equation~\eqref{eq:general-lower-master} yields \[ E[T] \ge \frac{f(a,A)^2}{4} E[T_A].\]
\end{proof}

%% file: si/proofs/double_star.tex
\section{Double Stars}

In this section, we examine the diversity times on double star in bd updating.
Double star has $N=2n+2$ vertices and consists of two stars $S_n$ and $S'_n$ where the centers are connected.

We say the configuration is \emph{neutral} if the red type occupies $S_n$ and the blue type occupies $S'_n$.
We say that the center of a star \emph{interacts} with the other star if it first spreads to the center of the other star and then, without being replaced, this center of the star claims a leaf.
The interaction happens in two active steps (where we count the active step as a reproduction of the first center and then the successful reproduction or death of the second center).

\begin{lemma}\label{lemma:small_interaction}
    The center of one star interacts with the other star with probability $\frac{n}{(n+1)^4}$.
\end{lemma}
\begin{proof}
    A center is selected with probability $\frac{1}{n+1}$ ($=\frac{2}{2n+2}$) and with probability $\frac{1}{n+1}$ it spreads to the other center.
    We consider only two possibilities as an active step, either the newly conquered center dies or reproduces to a leaf.
    The reproduction happens with probability $\frac{1}{n+1}\cdot \frac{n}{n+1}$, otherwise it dies.
    Multiplying the probabilities gives $\frac{n}{(n+1)^4}$.
\end{proof}

\subsection{Lower bound}
In the proof, we count only the number of steps in the neutral configuration.
It still gives us a bound of $\Omega(N^4)$.

\begin{lemma}\label{lem:co-jo-lb}
  The diversity time of a double star under bd updating is $\Omega(N^4)$.
\end{lemma}
\begin{proof}
  We will use \Cref{lemma:general_lower} starting in the neutral configuration
  so that $A$ is the set of vertices on the right star initially occupied by type $a=2$.
  In the neutral configuration, both types have the same probability of fixating.
  Therefore the fixation probability of type 2 in the neutral configuration is $f(a, A)=1/2$.
  By \Cref{lemma:general_lower}, the time until absorption $T$ satisfies
  \begin{equation}
    E[T] \geq f(a,A)^2 E[T_A] / 4 = E[T_A]/16 = \Omega(E[T_A]).
  \end{equation}
  Thus it suffices to find a time lower bound for the invincible process starting from the neutral configuration.
  In what follows, we analyze the time in the invincible process.
  
  By \Cref{lemma:small_interaction}, we know that the center of one star interacts with the other star with probability at most $1/n^3$.
  This means that the expected number of steps until type $2$ interacts with the other star starting from the neutral configuration is at least $\Omega(n^3)$.
  We will show that once type $2$ interacts with the other star, type $2$ has at most a $\calO(1/n)$ probability of fixating.
  With remaining probability, we return back to the neutral configuration.
  This means type $2$ will need to interact with the other star at least $\Omega(n)$ times in expectation before type $2$ fixates.
  Since the time between tries is $\Omega(n^3)$ in expectation, the total expected time is $\Omega(n^4)$ for the invincible process.

  We consider a process on a different graph with fixation probability at most $\calO(1/n)$
  and show that this is an upper bound for the probability type $2$ fixates in the other star in the invincible process between interactions.

  Consider a star with $n+1$ vertices where mutant (i.e. type $2$) have relative reproductive fitness $1+2/n$.
  We call this the \emph{improved} process.
  We show that for the improved process, we have a higher ratio between increasing the number of mutants and decreasing the number of mutants
  as opposed to the invincible process.
  
  For the improved process, the fixation probability with mutants at the center and a leaf is $\calO(1/n)$ from \cite{Broom2008,hadjichrysanthou2011evolutionary}.


  \paragraph{Center is occupied.}
  Suppose the center and $i$ leaves belong to type $2$ in the invincible process.
  We have a probability of increasing the number of type $2$ individuals $\frac{1}{n+2}\cdot\frac{n-i}{n+1}$.
  The probability of losing the center is $\frac{n-i}{n+2}$.
  The ratio is $\frac{1}{n+1}$.

  Now, suppose the center and $i$ leaves belong to mutants in the improved process.
  The probability of increasing the number of mutants is $\frac{1+\frac{2}{n}}{F}\frac{n-i}{n}$, where $F$ is the total fitness of all individuals.
  The probability of losing the center is $\frac{n-i}{F}$.
  The ratio is $\frac{1+\frac{2}{n}}{n}$.

  \paragraph{Center is not occupied.}
  Suppose $i$ leaves belong to type $2$ and the center belongs to type $1$ in the invincible process.
  We have a probability of claiming the center $\frac{i}{n+2} + \frac{1}{n+2}\cdot \frac{1}{n+1}$.
  The probability of losing a type $2$ individual is $\frac{1}{n+2}\frac{i}{n+1}$.
  The ratio is $n+1 + \frac{1}{i}$.

  Now, suppose $i$ leaves belong to mutants and the center belongs to the residents in the improved process.
  The probability of claiming the center is  $\frac{(1+\frac{2}{n})i}{F}$, where $F$ is the total fitness of all individuals.
  The probability of losing a mutant is $\frac{1}{F}\frac{i}{n}$.
  The ratio is $n+2$.

\paragraph{ }
  In both cases, we have a higher probability ratio of type $2$ increasing versus decreasing.
  This is what we wanted to show.
\end{proof}

\subsection{Upper bound}

\begin{lemma}\label{lemma:double_upper}
  The diversity time of a double star under bd updating is $\calO(N^4)$.
\end{lemma}
\begin{proof}
  First, we prove that all except two types disappear quickly.
  Then, we argue about the diversity time for only two types.

  \paragraph{Multiple types.}
  A double star consists of two $n$-leaf stars, $S_1$ and $S_2$, connected by an edge at their centers, $c_1$ and $c_2$, respectively.
  Suppose we start the process with $N$ types.
  Consider the vertices of $S_1$.
  Using a similar technique to that of \Cref{thm:bd-star-upper}, we show that $S_1$ becomes homogeneous in at most $O(n^3)$ steps in expectation.
  
  Let $\Pi$ be all partitions of the $n$ leaves of $S_1$.
  We define a potential function $\phi\colon\Pi\to \Real$
  that maps a given partition $\lambda$ of $\{1,\ldots,n\}$
  to the expected absorption time on a complete graph with $n$ vertices
  with initial configuration $\lambda$.
  
  We will track only the reproductions of individuals on the central vertex $c_1$ and the configuration in the leaves of the star $S_1$.
  Any time the center individual at location $c_1$ reproduces in $S_1$ (which happens with probability $(1/N) \cdot n/(n+1)$), four things could have happened in the previous round:
  \begin{enumerate}
      \item[(A)] an individual from a leaf of $S_1$ replaced the center individual $c_1$
      \item[(B)] the center individual $c_1$ reproduced onto a leaf of $S_1$,
      \item[(C)] the individual at the other center $c_2$ gave birth onto $c_1$, or
      \item[(D)] a death occurred in $S_2$.
  \end{enumerate}
  At each time step
  \begin{itemize}
      \item situation (A) has probability $p_A := n/N$ of occurring,
      \item situation (B) has probability $p_B := (1/N) \cdot n/(n+1)$ of occurring,
      \item situation (C) has probability $p_C := (1/N)\cdot 1/(n+1)$ of occurring, and
      \item situation (D) has probability $p_D := 1-p_A-p_B-p_C=(n/N) + (1/N)\cdot n/(n+1) + (1/N)\cdot 1/(n+1) = 1/2$ of occurring.
  \end{itemize}
  
  Now, for the configuration $\lambda\in\Pi$ on the leaves of the star, we have the following potential.
  \begin{equation*}
          \phi(\lambda) = \E{t^{\text{hit}}_\lambda} = n^2 - n - \sum_{i=1}^\ell\sum_{k=0}^{\lambda_i-1}\frac{(n+\lambda_i-2k)k}{n-k}.
  \end{equation*}
  Similar to the proof of \Cref{thm:bd-star-upper}, the largest change in the potential at any step is $n\log n$.
  
  Let $\lambda^{(i)}$ be the configuration of the leaves after $i$ steps of the process.
  Suppose the center reproduces for the $i$th step.
  Then
  \begin{itemize}
      \item with probability $p_A$, situation (A) occurs for the $(i-1)$th step.
      Then the potential changes by \[
          \phi(\lambda^{(i)})-\phi(\lambda^{(i-1)})
          =-1
      \]
      since this is one step in the process for the
      complete graph with self-loops.
      The expected absorption time (which is the potential function) will decrease by 1.
  
      \item
      with probability $p_B$, situation (B) occurs for the $(i-1)$th step.
      \begin{itemize}
          \item with probability $p_A$, situation (A) occurs for the $(i-2)$th step.
          Thus with probability at most $2/n$ the potential function has increased by at most $n\log n$ and with remaining probability has decreased by $1$ (see the details of \Cref{thm:bd-star-upper}.
          
          \item with probability $p_B$, situation (B) occurs for the $(i-2)$th step.
          By similar logic, the potential can only increase by $n\log n$ at each step.

          \item with probability $p_C$, situation (C) occurs for the $(i-2)$th step.
          Thus potential again can only increased by $n\log n$.

          \item with probability $p_D$, situation (D) occurs for the $(i-2)$th step and there is no change in the potential function.
      \end{itemize}

      \item with probability $p_C$, situation (C) occurs in the $(i-1)$th step and the potential can only increase by at most $n\log n$

      \item with probability $p_D$, situation (D) occurs in the $(i-1)$th step and the potential does not change.
  \end{itemize}
  
  Putting the pieces together, we get
  \begin{align*}
      E[\phi(\lambda^{(i)})-\phi(\lambda^{(i-1)})\mid \lambda^{(i-1)}]
      &\leq
      p_A\cdot(-1)  \\
      &+ p_B\cdot [p_A((2/n)\cdot n\log n + (1-2/n)\cdot (-1)) + (p_B+p_C)\cdot n\log n + p_D\cdot 0] \\
      &+ p_C\cdot n\log n \\
      &+ p_D\cdot 0 \\
      &\leq
      -p_A + p_A p_B\cdot (2\log n -1+2/n) + p_B^2\cdot n\log n + p_C (p_B+1)\cdot n\log n \\
      &=
      -\frac{n^3}{2(n+1)^3} 
      - \frac{5n^2}{4(n+1)^3}
      + \frac{5n^2\log n}{4(n+1)^3}
      + \frac{n\log n}{2(n+1)^3} \\
      &\leq -1/2 + o(1).
  \end{align*}
  From the Theorem 7 of \cite{kotzing2019first}, $S_1$ becomes homogeneous in $O(N^3)$ steps in expectation.
  
  Let $T_{S_1}$ ($T_{S_2}$) be the time it takes until $S_1$ ($S_2$) is homogeneous starting with $S_1$ ($S_2$) as heterogeneous.
  From above we have that $E[T_{S_1}] = O(N^3)$.
  By Markov's inequality, there exist some constants $c>0$ and $d\in (0,1)$ such that
  \begin{equation}
      P(T_{S_1}\geq cN^3) \leq E[T_{S_1}] / (cN^3) \leq d.
  \end{equation}
  Let $T_2$ be the time it takes until $S_1$ and $S_2$ are homogeneous starting with $S_1$ as heterogeneous.
  Let $T'$ be the time it takes until $S_2$ is homogeneous given that $S_1$ is homogeneous.
  Then for indicator random variable $I\{\cdot\}$ we have the system
  \begin{align*}
      T_2 &\leq T_{S_1} + T' \\
      T' &\leq I\{T_{S_2} < cN^3\} \cdot T_{S_2} +  I\{T_{S_2} \geq cN^3\} \cdot T_2.
  \end{align*}
  Taking expectations yields the system
  \begin{align*}
      E[T_2] &\leq O(N^3) + E[T'] \\
      E[T'] &\leq 1 \cdot cN^3 +  d \cdot E[T_2].
  \end{align*}
  This implies that $E[T_2] \leq O(N^3)$.

  \paragraph{Two types.}
  Having two types (type $1$ and type $2$) in the neutral configuration, we compute the expected time.
  From any configuration with more than two types, we reach the neutral configuration in $\calO(n^3)$ steps with constant probability.
  This result follows from the previous part of the proof for multiple types.
  
  Without loss of generality, suppose that type $2$ (on star $S_2$) invades a leaf of the star that type $1$ occupies ($S_1$).
  This happens with probability \[ p:=(1/N)(1/(n+1)) \cdot (1/N)\cdot (n/(n+1)) = \Theta(1/n^3) \] in two steps.
  
  Suppose no reproduction from $c_1$ to $c_2$ happens in $1/p = \calO(n^3)$ steps.
  This event occurs with constant positive probability $q$ by Markov's inequality.
  Thus $S_1$ is invaded without $S_2$ being invaded.
  The invader of $S_1$ has probability at least $\Omega(1/n)$ of conquering $S_1$.
  If one or both stars are invaded, in $\calO(n^3)$ steps on average, fixation or the neutral configuration is reached.
  But the neutral configuration is reached with probability $p\cdot \Omega(1/N)=\Omega(1/n^4)$ ($S_1$ invades $S_2$ and succeeds while $S_1$ is being invaded).
  This implies that the number of visits of all configurations that are not neutral configuration is at most $\calO(n^4)$.
  Moreover, the number of steps until at least one star is invaded from the neutral configuration is $\calO(n^3)$ in expectation.
  This is because from \Cref{lemma:small_interaction}, the probability of spreading between stars is at least $\Omega(1/n^3)$.

  With probability $1-q$, we restart our analysis with $O(n^3)$ steps taken.
  This gives a recurrence for the expected time $T$ until absorption of
  \[ T \leq (1-q)(T+\calO(n^3)) + q \cdot \calO(n^4)\]
  which has a solution of $T \leq \calO(N^4)$
  
\end{proof}

%% file: si/proofs/barbell.tex
\section{Barbells}
\subsection{db lower bound on the highest time}

A \emph{barbell} graph with $N=3n$ vertices consists of a left clique with $n$ vertices,
a path with $n$ vertices,
and a right clique with $n$.
We assume $n$ is even for simplicity.
One end of the path is connected to one vertex in the left clique.
The other end of the path is connected to one vertex in the right clique.
    
\begin{theorem}\label{thm:barbell_long}
    In the barbell graph, the diversity time is $\Omega(N^4)$.
\end{theorem}
\begin{proof}
    We decompose a barbell graph on $N=3n$ vertices
    as cliques $K_L$ and $K_R$ that each have a path $P_L$ (resp. $P_R$) of length $n/2$ connected to one of its nodes by an edge.
    Then $P_L$ and $P_R$ are connected by an edge at the endpoints of each of the paths (i.e. at the vertices that have unity degree).
    A configuration of types on a barbell graph is called \emph{neutral} if one type (type 1) occupies $V(K_L)\cup V(P_L)$
    and another type (type 2) occupies $V(K_R)\cup V(P_R)$.
    We prove a lower bound for when the process starts in a neutral configuration.
    We use \Cref{lemma:general_lower} from this configuration to prove the lower bound we desire.
    We will use $A=V(K_R)\cup V(P_R)$ as the invincible set of vertices and set type $a=2$.
    We have that the fixation probability of type $a=2$ is $f(a,A)=1/2$ by symmetry.
    Therefore by \Cref{lemma:general_lower},
    \[ E[T] \geq f(a,A)^2 E[T_A]/4 = E[T_A]/16 = \Omega(E[T_A]). \]
    Therefore it suffices to find a time lower bound for the invincible process starting from the neutral configuration.
    Below, we will analyze the time in the invincible process.

    We define four (out of many) possible configurations with two types:
    \begin{enumerate}
        \item[(I)] type 1 occupies exactly $V(K_L) \cup V(P_L)$,
        \item[(II)] type 1 occupies exactly $V(K_L)$,
        \item[(III)] type 1 occupies a proper non-empty subset of $V(K_L)$, and
        \item[(IV)] type 1 does not occupy any vertices.
    \end{enumerate}
    In order for the invincible process to absorb,
    we must pass through each of the above four configurations at least once.
    Let $T_I$, $T_{II}$, and $T_{III}$ be random variables for the number of steps it takes to reach configuration (IV) in the invincible process
    starting at the corresponding configurations.

    Firstly, we have \[ E[T_I]\geq \Omega(N^3) + E[T_{II}] \]
    from \Cref{thm:path}.
    We denote the vertex of $K_L$ connected to $P_L$ by and edge as the \emph{bridgehead}.
    
    In configuration (II), with high probability $(1-2/N)$ the configuration does not change since either the bridgehead or the node in $P_L$ connected to the bridgehead needs to be selected for death for the configuration to change.
    Losing the bridgehead to type 2 happens with probability at least $\Omega(1/N^2)$ since this could happen if immediately,
    the bridgehead is selected for death and the node in $P_L$ connected to the bridgehead is selected for birth.
    This is only two steps of the process.
    On the other hand, it is possible to back to configuration (I): the node in $P_L$ connected to the bridgehead
    is selected for death immediately and then the bridgehead is selected for birth.
    Then there is at least an $\Omega(1/N)$ chance by analyzing the fixation probability of an invader placed at the end of a path.
    This would take at least $\Omega(N)$ steps to achieve.
    Thus \[ E[T_{II}] \geq \Omega(1) + (1-2/N)\cdot E[T_{II}] + \Omega(1/N^2)\cdot E[T_{III}] + \Omega(1/N^2)\cdot (N+E[T_I]). \]
    Simplifying this expression gives
    \begin{align*}
    E[T_{II}] &\geq \Omega(N) + \Omega(1/N)\cdot E[T_{III}] + \Omega(1/N)\cdot E[T_I] \\
    &\geq \Omega(N) + \Omega(1/N)\cdot E[T_{III}] + \Omega(1/N)\cdot\left(\Omega(N^3) + E[T_{II}]\right) \\
    &\geq \Omega(N^3) + \Omega(E[T_{III}]).
    \end{align*}


    Consider an altered process such that after reaching configuration (III),
    no vertex in $P_L$ can be replaced (from death) until configuration (II) or (IV) is reached.
    This unilaterally decreases the fixation time of type 2.
    
    Now, we bound the probability of reaching configuration (IV) from configuration (III) without first reaching configuration (II) in this altered process.

    We examine the transition probabilities of a Markov Chain that tracks the number of type 2 individuals inside $K_L$.
    Then we examine the transition probabilities in a complete graph on $n$ vertices where one type has higher fitness (i.e. the modified scenario).
    We show that the ratio between increasing and decreasing the number of type 2 individuals is higher in this modified scenario for every configuration.
    This will mean that the fixation probability in this modified scenario is an upper bound for the altered process.

    \begin{itemize}[align=left]
    \item[\textbf{Invincible process.}] When there are $i$ type $2$ individuals, we have two possibilities for the probabilities.
    \begin{itemize}
    \item If the bridgehead is occupied by type $2$ individual, the probability that the number of type $2$ individuals increases is
    \[
        \frac{n-i}{n}\frac{i}{n-1}.
    \]
    The probability that the number of type $2$ individuals decreases is
    \[
        \frac{1}{n}\frac{n-i}{n} + \frac{i-1}{n}\frac{n-i}{n-1}\,.
    \]
    (Note that if the bridgehead dies, it has a smaller probability of being replaced by a type $1$ individual.)
    The ratio between the probability of increase and the probability of decrease is
    \begin{align*}
        \frac{\frac{n-i}{n}\frac{i}{n-1}}{\frac{1}{n}\frac{n-i}{n} + \frac{i-1}{n}\frac{n-i}{n-1}} = \frac{i}{\frac{n-1}{n} + i-1} = \frac{i}{i-\frac{1}{n}} \le 1 + \frac{1}{n-1}\,.
    \end{align*}

    \item If the bridgehead is occupied by type $1$ individual, the probability that the number of type $2$ individuals increase is
    \[
        \frac{1}{n}\frac{n-i-1}{n-1} + \frac{n-i-1}{n}\frac{i}{n}\,,
    \]
    since when the bridgehead dies, it can be replaced by the individual on the path.
    The probability that the number of type $2$ individuals decreases is
    \[
        \frac{i}{n}\frac{n-i}{n-1}\,.
    \]
    This gives us the ratio between the probability of increase and the probability of decrease
    \[
        \frac{\frac{1}{n}\frac{n-i-1}{n-1} + \frac{n-i-1}{n}\frac{i}{n}}{\frac{i}{n}\frac{n-i}{n-1}} = \frac{\frac{n-1}{n}(i+1) + (n-i-1)i}{i(n-i)} = \frac{(1-\frac{1}{n})(1+\frac{1}{i}) + n-i-1}{n-i} = 1 + \frac{\frac{1}{i} - \frac{1}{n} - \frac{1}{ni}}{n-i} \le 1 + \frac{1}{ni} \le 1 + \frac{1}{n}\,.
    \]
    \end{itemize}

    \item[\textbf{Altered process.}] Now, let us imagine the altered scenario where the type $2$ (later called mutants) has fitness $1 + \frac{2}{n-1}$.
    The ratio between increasing and decreasing the number of mutants in configuration $i$ is
    \[
        \frac{\frac{n-i}{n}\frac{i(1+\frac{2}{n-1})}{F-1}}{\frac{i}{n}\frac{n-i}{F-1-\frac{2}{n}}}\,,
    \]
    where $F$ is the sum of the fitnesses of all mutants in the graph.
    This gives the ratio between the probability of increase and the probability of decrease
    \[
        \frac{(1+\frac{2}{n-1})(F-1-\frac{2}{n})}{F-1} = 1 + \frac{2}{n-1} - \frac{(1+\frac{2}{n-1})\frac{2}{n-1}}{F-1} \le 1 + \frac{2}{n-1} - (1+\frac{2}{n-1})\frac{2}{(n-1)^2} \le 1 + \frac{1}{n-1}\,,
    \]
    where the last inequality holds for $n\ge3$.
    \end{itemize}

    Now, the ratio between increasing and decreasing the number of mutants is higher than in every configuration of the original case.
    That means the fixation probability in this altered scenario is an upper bound for the invincible process.
    From \S 6.2 of~\cite{nowak2006evolutionary}, the fixation probability in this case is $\calO(\frac{1}{n})$.
    This means that
    \begin{align*}
        E[T_{III}] \geq \Omega(1) + (1-\calO(1/N)) \cdot E[T_{II}].
    \end{align*}
    Simplifying yields
    \begin{align*}
        E[T_{III}] &\geq \Omega(1) + (1-\calO(1/N)) \cdot \left(\Omega(N^3) + \Omega(E[T_{III}])\right) \\
                   &\geq \Omega(N^4).
    \end{align*}

    Putting all of the pieces together, we get
    \begin{align*}
        E[T_I] \geq \Omega(N^4).
    \end{align*}
    This is what we wanted to prove. 
\end{proof}

%% file: si/proofs/upper_lower_bounds.tex
\section{Bounds for any graph}
In what follows, we use notation introduced in \S 1.1 of \cite{goldberg2024parameterised}.
We let $\AbsorptionTime_{\tau,f,X_t}(G)$
be the expected absorption time
of $G$ (under a specified dynamic)
starting at state $X_t$.
We let $\AbsorptionTime_{r=1}(G)$ be the maximum expected bd absorption time
over all possible initial mutant configurations
when the fitness function is constant (i.e. $|f(V)|=1$),
and there are only two types initially in the population
(i.e. both $|\tau| = 2$ and $X_0(V) = \tau$).
Recall that $|X_t(V)|$ denotes the number of types at time $t$.
(see~\Cref{section:model}).

\subsection{Upper bound on expected absorption time under bd updating}
\begin{lemma}\label{lemma:ceiling-square}
   Suppose $k\geq 2$ is an integer.
   Then $\lceil k/2\rceil^2 \leq k^2/2$.
\end{lemma}
\begin{proof}
   If $k$ is even then $\lceil k/2\rceil^2 = k^2/4 \leq k^2/2$.
   If $k$ is odd, it must be the case that $k\geq 3>(\sqrt2-1)^{-1}$.
   Thus $1 + 1/k \leq \sqrt 2$.
   Taking the logarithm of both sides gives us that
   $\log_2 (k+1) - \log_2 k \leq 1/2$.
   By rearranging terms and multiplying both sides by $2$, we get
   $2\log_2 (k+1) - 1 \leq 2\log_2 k$ which means
   \begin{equation}
       (k+1)^2/4 \leq k^2/2.
   \end{equation}
   Finally, we notice that when $k$ is odd, $\lceil k/2\rceil^2 = (k+1)^2/4$.
\end{proof}

In the following theorem,
we show that the expected absorption time in the process with $|X_t(\Vertices)|$ types is at most $\log_2 |X_t(\Vertices)|$ longer than the process with only two types.
\begin{theorem}\label{theorem:neutral-upper-bound}
    Let $\Graph=(\Vertices,\Edges)$ be a graph on $N$ vertices.
    Let $|f(\Vertices)|=1$.
    Then for bd updating, $\AbsorptionTime_{\tau, f, X_t}(\Graph)
    \leq 2\cdot\AbsorptionTime_{r=1}(\Graph)\cdot \log_2|X_t(\Vertices)|$
    for all $t\geq 0$.
\end{theorem}
\begin{proof}
    We will map the multi-type Moran process
    to a two type Moran process.
    Clearly if $|X_t(\Vertices)| = 1$ the process has absorbed and thus its expected absorption time is upper bounded by $2\cdot\AbsorptionTime_{r=1}(\Graph)\cdot \log_2|X_t(\Vertices)|$.
    Suppose $\AbsorptionTime_{\tau,f,X_t}(\Graph)\leq 2\cdot\AbsorptionTime_{r=1}(\Graph)\cdot \log_2 k$ for all $X_t\in\Omega$ such that $|X_t(\Vertices)| < k$.
    Now consider the case when $|X_t(\Vertices)| = k > 1$.
    Let $\theta_{X_t}\colon X_t(\Vertices)\to\Set{\texttt A, \texttt B}$
    such that 
    there is a subset of types $S\subseteq X_t(\Vertices)$ of size $\lceil X_t(\Vertices)/2 \rceil$
    with $\theta_{X_t}(S)=\Set{\texttt A}$
    and $\theta_{X_t}(X_t(\Vertices)\setminus S) = \Set{\texttt B}$.
    This is always possible since $X_t(\Vertices)$ is finite 
    and thus under an arbitrary ordering of the elements of the set, we map the first $\lceil X_t(\Vertices)/2 \rceil$ elements to $\texttt A$ and the rest to $\texttt B$.
    Then we run the multi-type Moran process on
    $(\tau, f, \theta_{X_t}\circ X_t)$ until absorption
    which takes $T_k$ steps. 
    Notice that since $|(\theta_{X_t}\circ X_t)(\Vertices)|\leq 2$, this is the Moran process
    on a graph with two types under neutral evolution.
    When this process absorbs, it must be the case that 
    $|(\theta_{X_t}\circ X_{t+T})(\Vertices)| = 1$
    which in turn means
    \begin{equation}
        |X_{t+T_k}(\Vertices)| \leq \max\Set{|\theta_{X_t}(S)|, |\theta_{X_t}(\Vertices\setminus S)|} \leq \lceil |X_t(\Vertices)|/2 \rceil = \lceil k/2 \rceil.
    \end{equation}
    Thus by linearity of expectation we get 
    \begin{equation}
        \AbsorptionTime_{\tau,f,X_t}(\Graph)
        \leq 
        \E{T_k}
        + 
        \E{\AbsorptionTime_{\tau,f,X_{t+T_k}}(\Graph)}.
    \end{equation}
    Since $|X_{t+T_k}(\Vertices)| \leq \lceil k/2 \rceil < k$ for $k>1$, the inductive hypothesis gives us
    \begin{equation}
        \E{\AbsorptionTime_{\tau,f,X_{t+T_k}}(\Graph)}
        \leq 
        2\cdot\AbsorptionTime_{r=1}(\Graph)\cdot \log_2\lceil k/2\rceil.
    \end{equation}
    Finally, we get
    \begin{align}
        \AbsorptionTime_{\tau,f,X_t}(\Graph)
        &\leq \AbsorptionTime_{r=1}(\Graph) + 2\cdot\AbsorptionTime_{r=1}(\Graph)\cdot \log_2\lceil k/2\rceil \\
        &= 2\cdot\AbsorptionTime_{r=1}(\Graph)\cdot (1+\log_2\lceil k/2\rceil^2) \\
        &= \AbsorptionTime_{r=1}(\Graph)\cdot\log_22\lceil k/2\rceil^2 \\
        &\leq 2\cdot \AbsorptionTime_{r=1}(\Graph)\cdot\log_2 k
    \end{align}
    where the last step follows from~\cref{lemma:ceiling-square}.
\end{proof}

\begin{corollary}
    Let $\Graph=(\Vertices,\Edges)$ be a graph on $N$ vertices.
    Let $|f(\Vertices)|=1$.
    Then for bd updating, $\AbsorptionTime_{\tau, f, X_t}(\Graph)
    \leq
    2N^6\log_2|X_t(\Vertices)|$ for all $t\geq 0$.
\end{corollary}
\begin{proof}
    By Corallary 12(i) of~\cite{diaz2014approximating},
    we have $\AbsorptionTime_{r=1}(\Graph)\leq\phi_\Graph(\Vertices)^2 N^4\leq N^6$.
    The result follows from~\cref{theorem:neutral-upper-bound}.
\end{proof}

\subsection{Lower bound on expected absorption time under bd updating}

\begin{theorem}\label{theorem:neutral-lower-bound}
    Let $\Graph=(\Vertices,\Edges)$ be a graph on $N$ vertices.
    Let $|f(\Vertices)|=1$.
    Fix $t \geq 0$ and let $k := |X_t(\Vertices)|>1$.
    Then
    \begin{equation}
    \AbsorptionTime_{\tau, f, X_t}(\Graph)
    \geq N\cdot(1-1/k)\cdot H_{N\cdot(1-1/k)}
    \geq N\cdot(1-1/k)\ln(N\cdot(1-1/k))
    \end{equation}
    where $H_m := \sum_{i=1}^m 1/i$ is the $m^{\text{th}}$ harmonic number.
\end{theorem}
\begin{proof}
    In order for absorption to occur on a graph with $k$ types,
    it must be the case that at least $k-1$ distinct vertices are selected for death since absorption only occurs when one type remains.
    Thus a lower bound on the absorption time is the expected amount of time for $k-1$ vertices to be selected for death.
    Notice that the probability that a vertex $u\in\Vertices$ is selected for death is independent of $f$ and $t$; let $p_u$ be this probability.
    We have $p_u = T(u)/N$ where
    \begin{equation}
        T(u) := \sum_{v\in\Gamma(u)}\frac{1}{\deg(v)}
    \end{equation}
    is the \textit{temperature} of vertex $u\in\Vertices$.
    
    We will make use of the coupon collector problem (see here~\cite{Flajolet1992}).
    For a probability vector $\pi\in[0,1]^m$, we denote
    $\mathsf{CC}_{\pi}$ as the number of coupon draws 
    until each distinct type of $m$ coupons has been drawn at least once,
    where $\pi_i$ is the probability that the $i^{\text{th}}$ type of coupon is drawn.
    By considering deaths at vertices as coupons, we have that 
    \begin{equation}
         \AbsorptionTime_{\tau, f, X_t}(\Graph)
         \geq 
         \min_{i\in X_t(\Vertices)}{\mathsf{CC}_{\rho^{(-i)}}}
    \end{equation}
    where $\rho^{(-i)}$ is the $|\Vertices\setminus X_t^{-1}(i)|$-dimensional probability vector such that
    \begin{equation}
        \rho^{(-i)}_u \propto p_u
    \end{equation}
    for each $u\in\Vertices\setminus X_t^{-1}(i)$;
    this is because (1) coupon draws are independent, and 
    (2) the number of draws until a subset $S$ of types of coupons
    have each been drawn
    is at least
    the number of draws \textit{restricted to coupons whose types are in $S$} until the subset $S$ of types of coupons
    have each been drawn.
    From Corollary 4.2 of~\cite{Flajolet1992}, we have
    \begin{equation}
        \mathsf{CC}_{\pi} = \sum_{q=0}^{m-1}(-1)^{m-1-q}\sum_{\substack{J\subseteq [N]\\ |J|=q}}\left(1-\sum_{j\in J}\pi_j\right)^{-1}
    \end{equation}
    and $\mathsf{CC}_\pi = mH_{m}$ when $\pi$ is the uniform
    distribution.
    By~\cite{clevenson1991majorization}, $\mathsf{CC}_{\rho^{(-i)}}$ is Schur-convex and thus is minimized when $\rho^{(-i)}$ is uniform.
    Thus we have 
    \begin{equation}
        \min_{i\in X_t(\Vertices)}{\mathsf{CC}_{\rho^{(-i)}}}
        \geq
        \min_{i\in X_t(\Vertices)}|\Vertices\setminus X_t^{-1}(i)|\cdot H_{|\Vertices\setminus X_t^{-1}(i)|}.
    \end{equation}
    Since $\sum_{i\in X_t(\Vertices)} |X_t^{-1}(i)| = N$
    and $|X_t^{-1}(i)|\geq 1$ for each $i\in X_t(\Vertices)$,
    we must have $\max_{i\in X_t(\Vertices)} \geq N/k$.
    This gives us
    \begin{equation}
        |\Vertices\setminus X_t^{-1}(i)| \geq N-\max_{j\in X_t(\Vertices)}
        |X_t^{-1}(j)|
        \geq N\cdot(1-1/k)
    \end{equation}
    for each $i\in X_t(\Vertices)$.
    Finally, this means that 
    \begin{equation}
        \min_{i\in X_t(\Vertices)}|\Vertices\setminus X_t^{-1}(i)|\cdot H_{|\Vertices\setminus X_t^{-1}(i)|}
        \geq N\cdot (1-1/k)\cdot H_{N\cdot (1-1/k)}.
    \end{equation}
    It is known that $H_m \geq \ln m$ for every $m\geq 1$.
\end{proof}

\subsection{Upper bound on expected absorption time under db updating}

For a graph $G=(V,E)$, let $M:=|E|$ be the number of edges,
let $\deg_{\min} := \min{\deg(u)\mid u\in V}$ be the minimum degree of the vertices, and for $S\subseteq V$ let $\deg(S) := \sum_{u\in S}\deg(u)$.
Let $\Phi$ be the \emph{conductance} of $G$, defined as
\begin{equation}
   \Phi := \min_{S\subseteq V} \frac{\#\{(u,v)\in E \mid u\in S \text{ and } v\not\in S\}}{\min\{\deg(S),\deg(V\setminus S)\}}
\end{equation}
where $0/0 = \infty$.

\begin{theorem}\label{thm:db-highest}
    The diversity time of $G$ under db updating is upper bounded by
    \begin{equation}
        \mathcal O\left(\frac{MN\log N}{\Phi\cdot \deg_{\min}}\right).
    \end{equation}
\end{theorem}
\begin{proof}
    See Theorem 2, Example 11 of \cite{cooper2016linear}
    for the solution when there are only two initial types.
    When there are $N$ initial types,
    we use the recursive reduction from \Cref{theorem:neutral-upper-bound} which adds a multiplicative logarithm factor based on the number of initial types.
\end{proof}

\begin{corollary}
    The diversity time of $G$ under db updating is upper bounded by $\mathcal O(M^2N\log N)$,
    which is at most $\mathcal O(N^5\log N)$.
\end{corollary}
\begin{proof}
    The result follows from \Cref{thm:db-highest} since the conductance of a connected graph is at least $\Omega(1/M)$ and there are at most $\mathcal O(N^2)$
    edges in a graph.
\end{proof}

\subsection{Lower bound on expected absorption time under db updating}

\begin{theorem}
    The diversity time of $G$ under db updating is lower bounded by $\Omega(N\log N)$.
\end{theorem}
\begin{proof}
    The proof is similar to that of \Cref{theorem:neutral-lower-bound} though is simpler since the probability
    a death occurs at a vertex is exactly $1/N$.
    Thus the lower bound is precisely the solution to the classic coupon collector problem~\cite{Flajolet1992}.
\end{proof}

%% file: si/proofs/directed_graph.tex
\section{Directed graphs}

First, we prove a general lemma about the upper bound on the diversity time on directed graph and then, we describe a contracting star: a graph that achieves that upper bound.

\begin{lemma}\label{lem:dir_upper}
    For all strongly connected directed graphs, the diversity time is at most
    $2^{\calO(N\log N)}$.
\end{lemma}
\begin{proof}
    In any configuration, we have a probability of at least $\frac{1}{N^2}$ that type A replaces type B.
    In any configuration, there are at most $N-1$ individuals of type B (other than A).
    That means, after $N-1$ steps, the probability that the process is finished is at least $\left(\frac{1}{N^2}\right)^{N-1}$.
    That means, we can bound the diversity time as
    \begin{align*}
        T_N &\le N-1 + \left(1-\left(\frac{1}{N^2}\right)^{N-1}\right) T_N.\\
    \end{align*}
    Rearranging, we get 
    \begin{align*}
        T_N \leq N^{2(N-1)}\cdot (N-1) \leq N^{2N}.
    \end{align*}
    That means the maximal expected time from any configuration is $2^{\calO(N\log N)}$.
\end{proof}

\subsection{Contracting stars}
A \emph{contracting star} graph consists of $k$ blades and one central vertex.
Every blade consists of $n$ vertices indexed from $1$ to $n$.
For each blade,
\begin{itemize}
    \item vertex $i$ is bi-directionally connected to vertex $i+1$ for each $i=1,\ldots,n-1$;
    \item there is an edge from vertex $i$ to the central vertex for each $i=1,\ldots,n$;
    \item there is an edge from vertex $j$ to vertex $i$ for each $1 \leq i < j \leq n$;
    \item there an edge from the central vertex to vertex $1$.
\end{itemize}
The contracting star with $k$ blades and $n$ blade length has $N = kn+1$ total vertices.

\begin{lemma}\label{lem:contracting_star}
    The diversity time for $k$ types on a contracting star with $k$ blades is $2^{\Omega(n \log n)}/k$.
\end{lemma}
\begin{proof}
    We lower bound the expected time until a vertex with index $n$ is replaced in a blade.
    Since we treat all blades independently, this is also a lower bound on the diversity time. 
    We suppose the blade is occupied by type $1$ while the type $2$ invades.
    To help the type $2$ individuals, we suppose that all individuals of type $2$ are invincible with the exception of the individual occupying the vertex with the highest index.
    
    Let us look at the type $2$ vertex with the highest index $i$.

    \begin{itemize}
        \item Under bd updating,
        type $2$ spreads to index $i$ with probability $\frac{1}{N}\frac{1}{i+1}$.
        But it is killed with probability $\frac{1}{N}\sum_{j > i}^{n} \frac{1}{j+1}$.
    
        For $\frac{1}{4}n \le i \le \frac{1}{2}n$, we have the probability of type $2$ to be killed
        \[
            \frac{1}{N}\frac{1}{n}\sum_{j > i}^{n} \frac{1}{j+1} > \frac{1}{N}\sum_{j > i}^{2i} \frac{1}{2i+2} \ge \frac{1}{3N}\,.
        \]
        So the ratio between reproduction and killing is at most $\frac{12}{n}$, which means the type $2$ is $n/12$ times more likely to die than to reproduce.
        That gives the expected time at least $(n/12)^{1/4 n}$ for one blade.
    
        \item Under db updating,
        type $2$ spreads to index $i$ with probability $\frac{1}{N}\frac{1}{n-(i-1)}$.
        
        But it is killed with probability $\frac{1}{N}\frac{n-(i-1)-1}{n-(i-1)}$
    
        For $\frac{1}{4}n \le i \le \frac{1}{2}n$, we have the probability of type $2$ to be killed is at least \[
            \frac{1}{N}\frac{n-(i-1)-1}{n-(i-1)}
            \geq \frac1N\left(1-\frac{1}{n-(n/4-1)}\right)
            \geq 1/N
        \]
        So the ratio between reproduction and killing is at most $2/n$, which means the type $2$ is $n/2$ times more likely to die than to reproduce.
        That gives the expected time at least $(n/2)^{1/4 n}$ for one blade.
    \end{itemize}

    Since the processes are independent in blades, we have the total diversity time $2^{\Omega(n\log n)}/k$.
    
\end{proof}

\begin{corollary}
    The diversity time on a contracting star with constant number of blades is $2^{\Theta(n\log n)}$.
\end{corollary}